\documentclass[10pt,onecolumn,draftcls]{IEEEtran}
\usepackage{verbatim}
\usepackage{amsfonts,amsmath,mathrsfs,amssymb,amsbsy}
\usepackage[final]{graphicx}
\usepackage{times,cite}
\begin{document}
\newtheorem{theorem}{Theorem}
\newtheorem{corrly}{Corollary}
\newtheorem{prop}{Proposition}
\newtheorem{lemma}{Lemma}
\newtheorem{definition}{Definition}
\newtheorem{remark}{Remark}
\newtheorem{claim}{Claim}
\newcommand{\entropy}[1]{\mathsf{h}\left(#1\right)}
\newcommand{\logfn}[1]{\log\left(#1\right)}
\newcommand{\twobyn}{\frac{2}{n}}
\newcommand{\nbytwo}{\frac{n}{2}}
\newcommand{\chanmatrix}{\text{H}}
\newcommand{\prob}{\textsf{P}}
\newcommand{\tr}[1]{\text{tr}\left(#1\right)}
\newcommand{\cov}[1]{\text{Cov}\left(#1\right)}
\newcommand{\inv}[1]{\left(#1\right)^{-1}}
\newcommand{\expect}[1]{\mathbb{E}\left[#1\right]}
\newcommand{\Csum}{\mathcal{C}_{\text{sum}}^{\text{IC}}}
\newcommand{\CsumIC}{\mathcal{C}_{\text{sum}}^{\text{IC}}}
\newcommand{\CsumGAIC}{\mathcal{C}_{\text{sum}}^{\text{GA-IC}}}
\newcommand{\snr}{P}
\newcommand{\BCbound}{\text{broadcast channel outer bound}}
\newcommand{\ZCbound}{\text{Z-channel outer bound}}
\newcommand{\muser}{$M$--user }
\newcommand{\gic}{Gaussian interference channel }
\newcommand{\onemany}{one-to-many }
\newcommand{\manyone}{many-to-one }
\newcommand{\RIC}{R^{\textrm{IC}}_{\textrm{TIN}}}
\newcommand{\RGAIC}{R^{\textrm{GA-IC}}_{\textrm{TIN}}}
\newcommand{\RDIFF}{R^{\textrm{DIFF}}_{\textrm{TIN}}}
\newcommand{\be}{\[\begin{split}}
\newcommand{\ee}{\end{split}\]}
\newcommand{\ud}{\underline{d}}
\newcommand{\uc}{\underline{c}}
\newcommand{\ucp}{\underline{c}_{\perp}}
\newcommand{\ub}{\underline{b}}
\newcommand{\vv}{\underline{v}}
\newcommand{\vu}{\underline{u}}
\newcommand{\udt}{\underline{d}}
\newcommand{\uct}{\underline{c}^{\top}}
\newcommand{\ucpt}{\underline{c}_{\perp}^{\top}}
\newcommand{\ubt}{\underline{b}^{\top}}
\newcommand{\vvt}{\underline{v}^{\top}}
\newcommand{\urho}{\underline{\rho}}
\newcommand{\ubd}{\ub^{\top}\ud}
\newcommand{\ucv}{\uc^{\top}\vv}
\newcommand{\ubc}{\ub^{\top}\uc}
\newcommand{\ucd}{\uc^{\top}\ud}
\newcommand{\ucpd}{\uc_{\perp}^{\top}\ud}
\newcommand{\ct}{\cos\theta}
\newcommand{\st}{\sin\theta}
\newcommand{\cst}{\cos^2\theta}
\newcommand{\sst}{\sin^2\theta}
\newcommand{\uy}{\underline{Y}}
\newcommand{\uyt}{\underline{\tilde{Y}}}
\newcommand{\ux}{\underline{X}}
\newcommand{\vx}{\underline{x}}
\newcommand{\uxt}{\tilde{X}}
\newcommand{\uxh}{\underline{\hat{X}}}
\newcommand{\uz}{\underline{Z}}
\newcommand{\uzt}{\underline{\tilde{Z}}}
\newcommand{\uzh}{\underline{\hat{Z}}}
\newcommand{\us}{\underline{S}}
\newcommand{\hs}{\hat{S}}
\newcommand{\ush}{\underline{\hat{X}}}
\newcommand{\ust}{\underline{\tilde{S}}}
\newcommand{\uh}{\underline{h}}
\newcommand{\uv}{\underline{V}}
\newcommand{\uvt}{\underline{\tilde{V}}}
\newcommand{\uw}{\underline{W}}
\newcommand{\uzero}{\underline{0}}
\newcommand{\bbH}{\mathbb{H}}
\newcommand{\bbA}{\mathbb{A}}
\newcommand{\bbQ}{\mathbb{Q}}
\newcommand{\bbC}{\mathbb{C}}
\newcommand{\SigmaGenie}{\Sigma}
\newcommand{\sigmaGenie}{\eta}
\newcommand{\lambdaMax}{\lambda_{\textrm{max}}}
\newcommand{\heff}{\tilde{h}}
\newcommand{\optbf}{\underline{b}}
\newcommand{\Q}{\hat{Q}}
\newcommand{\inr}{\textrm{INR}}

\title{Sum Capacity of MIMO Interference Channels in the Low Interference Regime}
\author{{\large{V. Sreekanth Annapureddy, {\em Student Member, IEEE,} and Venugopal V. Veeravalli, {\em Fellow, IEEE}}}
\thanks{The authors are with the Coordinated Science Laboratory and the
Department of Electrical and Computer Engineering,
University of Illinois at Urbana-Champaign, Urbana, IL 61801 USA (e-mail: sreekanthav@gmail.com, vvv@illinois.edu).}
\thanks{This research was supported in part by the NSF award CNS-0831670, through the University of Illinois, and grants from Intel and Texas Instruments.}
\thanks{This paper was presented in part at the Asilomar Conference on Systems, Signals and Computers, Pacific Grove, CA, November 2008, and the Information Theory and Applications (ITA) workshop, UCSD, San Diego CA, January 2009.}
}
\maketitle
\begin{abstract}
Using Gaussian inputs and treating interference as noise at the receivers has recently been shown to be sum capacity achieving for the two-user single-input single-output (SISO) Gaussian interference channel in a low interference regime, where the interference levels are below certain thresholds. In this paper, such a  low interference regime is characterized for multiple-input multiple-output (MIMO) Gaussian interference channels. Conditions are provided on the direct and cross channel gain matrices  under which using Gaussian inputs and treating interference as noise at the receivers is sum capacity achieving. For the special cases of the symmetric  multiple-input single-output  (MISO)  and single-input multiple-output (SIMO) Gaussian interference channels,  more explicit expressions for the low interference regime are derived. In particular, the threshold on the interference levels that characterize low interference regime is related to the input SNR and the angle between the direct and cross channel gain vectors.  It is shown that the low interference regime can be quite significant for MIMO interference channels, with the low interference threshold being at least as large as the sine of the angle between the direct and cross channel gain vectors for the MISO and SIMO cases.
\end{abstract}
\section{Introduction}
Breaking the interference barrier is an important step in achieving higher throughput in wireless networks. Towards this end, there has been a renewed interest in information-theoretic studies of interference networks in recent years. A canonical problem for such an  information-theoretic analysis is the two-user single-input single-output (SISO) Gaussian interference channel. This problem was first studied more than thirty years ago, and the capacity region was determined in the strong (and very strong) interference regimes, where the interference-to-noise ratio (INR) is larger than the signal-to-noise ratio (SNR)  at each receiver \cite{Carleial1975,HK1981, Sato1981}.

Establishing the capacity region in the more commonly encountered weak interference regime, where the INR is smaller than the SNR, is still mostly an open problem.  Nevertheless, the best known inner bound on the capacity region proposed by Han and Kobayashi  \cite{HK1981} was recently shown to be within one bit of the capacity region \cite{OneBit2008}. In the Han-Kobayashi scheme, the users split their messages into private and common messages, and each user jointly decodes its own message and the common message of the interfering user. This is in general a sophisticated scheme, requiring  multi-user encoders and decoders and coordination between the users. While such techniques are promising and being implemented in advanced systems, the traditional way to combat the interference is to treat interference as noise when the interference is weak, and to orthogonalize the users in time or frequency when the interference is moderate. Interestingly, treating interference as noise with Gaussian inputs was shown to be capacity achieving in a low (very weak) interference regime  \cite{Shang-Kramer-Chen-IT-2008,Motahari-Khandani-IT-2008,Annapureddy-Veeravalli-IT-2008}, which is the counterpart of the very strong interference regime \cite{Carleial1975}. Our goal in this paper is to characterize such a low interference regime for multiple-input multiple-output (MIMO) Gaussian interference channels. 

\begin{figure}[t]
\centering
\includegraphics[width=3in]{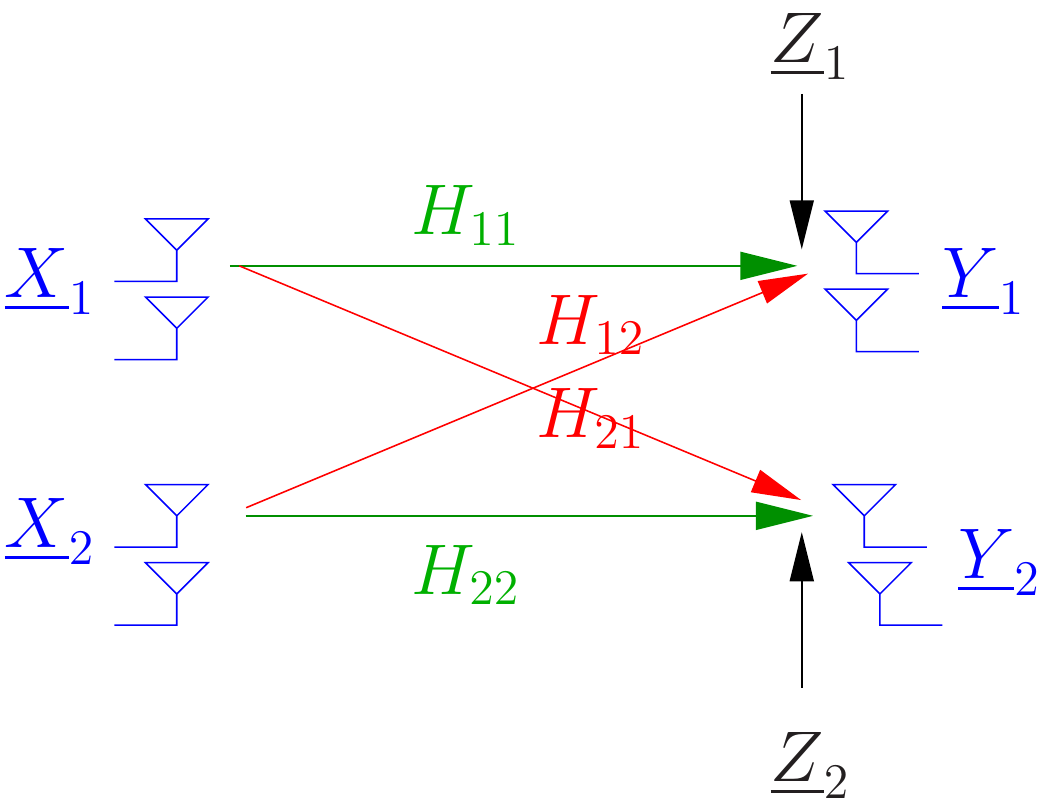}
\caption{Two-user MIMO Gaussian interference channel}
\label{fig:mimo_ic}
\end{figure}

The channel of interest is described in Figure~\ref{fig:mimo_ic}. There has been some previous work on studying inner and outer bounds on the capacity region of this channel \cite{Vishwanath-Jafar-04,Shang-Chen-Gans-06,TelatarTse2007,Shang-et-al-Allerton2008,shang_capacity_2009}. In particular, Teletar and Tse \cite{TelatarTse2007} showed that an appropriate extension of the SISO Han-Kobayashi inner bound is within a finite number (equal to the number of receive antennas) of bits of the capacity region. The sum capacity of the MIMO interference channel was analyzed previously by Shang et al \cite{Shang-et-al-Allerton2008,shang_capacity_2009}, where they showed that using Gaussian inputs and treating interference as noise is sum capacity achieving if the channel matrices satisfy a low interference regime condition for all input covariance matrices satisfying certain average transmit power constraints. For the special case of multiple-input single-output (MISO) interference channel, this result is not useful because the required low interference regime condition can be satisfied only if the input covariance matrices are unit-rank, but the result in \cite{Shang-et-al-Allerton2008} requires the condition to be satisfied for all input covariance matrices satisfying the average transmit power constraints. Bernd et. al. studied the low interference regime of MISO interference channel under the assumption that single-mode beamforming is applied at the transmitters \cite{Bernd-etal-Asilomar2008}.

For the general MIMO interference channel, we show that it is sufficient for the low interference regime condition to be satisfied only for the optimal input covariance matrices (those that maximize the achievable sum rate assuming Gaussian inputs and treating interference as noise) if they are full rank. For the MISO interference channel, this result cannot be exploited to establish a low interference regime because the optimal input covariance matrices are unit-rank, i.e., transmit beamforming maximizes the achievable sum-rate for the MISO interference channel \cite{shang_multi-user_2009}. We analyze the special case of symmetric MISO interference channel explicitly and derive a simple expression that characterizes the low interference regime. In particular, we relate the threshold on the interference levels that characterize low interference regime to the input SNR and the angle between the direct and cross channel gain vectors. We show that the same low interference regime holds for the dual single-input multiple-output (SIMO) interference channel also.

We establish that the low interference regime can be quite significant for MIMO interference channels, with the low interference threshold being at least as large as the sine of the angle between the direct and cross channel gain vectors for the MISO and SIMO cases.

\subsection{Notation and Organization}
We use underlined letters to denote vectors, and superscripts to denote sequences in time. For example, we use $X$ to denote a  scalar, $\ux$ to denote a vector, and $X^n$ and $\ux^n$ to denote sequences of length $n$ of scalars and vectors, respectively. We use $\cov{X}$ to denote the variance of a random variable $X$, and $\cov{X|Y}$ denote the minimum mean square error in estimating the random variable $X$ from the random variable $Y$, with similar notation for random vectors. We use $\mathcal {N} (\mu,\sigma^2)$ to denote the Gaussian distribution with mean $\mu$ and variance $\sigma^2$, and $\mathcal {N} (\underline{\mu},\Sigma)$ to denote the Gaussian vector distribution with mean $\underline{\mu}$ and covariance matrix $\Sigma$. We use $\mathsf{h}(.)$ to denote the differential entropy of a continuous random variable or vector and $I(.;.)$ to denote the mutual information. For real symmetric matrices $A$ and $B$, we use $A \succeq B$ to denote that $A-B$ is positive semidefinite.

The rest of the paper is organized as follows. In Section~\ref{sec:model}, we introduce the model for MIMO interference channel that we study. In Section~\ref{sec:mimo}, we present the results on the general MIMO interference channel. In Sections~\ref{sec:miso} and \ref{sec:simo}, we characterize the low interference regime for MISO and SIMO interference channels, respectively. In Section~\ref{sec:discussion}, we discuss some properties of the low interference regime. In Section~\ref{sec:concl}, we provide some concluding remarks.
\section{Problem Statement and Summary of Contributions} \label{sec:model}
Consider the two-user MIMO Gaussian interference channel (Figure~\ref{fig:mimo_ic}):
\begin{equation} \label{eq:mimo-ic}
\begin{split}
\uy_{1} = H_{11}\ux_{1} + H_{12}\ux_{2} + \uz_{1} \\
\uy_{2} = H_{21}\ux_{1} + H_{22}\ux_{2} + \uz_{2}
\end{split}
\end{equation}
with inputs $\ux_{1},\ux_{2}$ and the corresponding outputs $\uy_{1},\uy_{2}$. The channel gain matrices $\{H_{ij}\}$ are arbitrary but fixed and real. The receiver noise terms $\uz_{1} \sim \mathcal{N}(0,I)$ and $\uz_{2} \sim \mathcal{N}(0,I)$,  are assumed to be independent and identically distributed (i.i.d.) in time. The average transmit power constraints on users $1$ and $2$ are $P_{1}$ and $P_{2}$, respectively. Using $\ux_{it}$ to denote the channel input of the user $i$ at time $t$, the average transmit power constraints can be written as
\[
\frac{1}{n}\sum_{t=1}^{n} \expect{\ux_{it}\ux_{it}^{\top}} \in \mathcal{Q}_{i},
\]
where
\begin{equation} \label{eq:trace-constraint}
\mathcal{Q}_{i} = \{ Q_{i}: Q_{i} \succeq 0,  \tr{Q_{i}} \leq P_{i} \}.
\end{equation}We are interested in characterizing the low interference regime where using Gaussian inputs and treating interference as noise at the receivers achieves the sum capacity.
\subsection{Achievable sum-rate} \label{sec:mimoa}
Let $\RIC(Q_1,Q_2)$ denote the sum-rate achieved by  using Gaussian inputs with input covariance matrix $Q_i$ at transmitter $i$ and treating interference as noise at the receivers:
\[
\RIC(Q_1,Q_2) = \sum_{i=1}^2 I(\ux_{iG};\uy_{iG}) \\
\]
where the subscript $G$ on the inputs and outputs is used to indicate that Gaussian inputs are used. The input covariance matrices used should be obvious from the context throughout the paper.

The following result holds since any achievable sum rate is a lower bound on the sum capacity.
\begin{lemma} \label{lemma:lowerBound}
The sum capacity $\Csum$ of the MIMO interference channel \eqref{eq:mimo-ic} is lower bounded by
\begin{equation} \label{eq:max-ric}
\Csum \geq \RIC(Q_1^*,Q_2^*)
\end{equation}
where $(Q_1^*,Q_2^*)$ are optimal covariance matrices:
\begin{equation} \label{eq:optQ-mimo}
(Q_1^*,Q_2^*) \in \arg\max_{Q_{i} \in \mathcal{Q}_{i}} \RIC(Q_1,Q_2).
\end{equation}
\end{lemma}
\medskip
In the rest of the paper, we investigate when the inequality in Lemma~\ref{lemma:lowerBound} is achieved with equality, i.e., when treating interference as noise achieves the sum capacity. The optimization over the input covariance matrices is the new feature that makes the problem more difficult than its SISO counterpart, since little is known about the structure of the optimal covariance matrices $(Q_1^*,Q_2^*)$ for general MIMO interference channels. For MISO interference channels, it was shown in \cite{shang_multi-user_2009} that the optimal covariance matrices are unit rank; this fact is exploited in Section~\ref{sec:miso} to derive explicit conditions for the low interference regime.
\subsection{Genie-aided outer bound}
The outer bound is based on a genie giving side information to the receivers. Let $\us_{i}$ denote the side information given to the receiver $i$:
\begin{equation} \label{eq:genie-mimo}
\begin{split}
\us_1 = H_{21}\ux_1 + \uw_1 \\
\us_2 = H_{12}\ux_2 + \uw_2
\end{split}
\end{equation}
Let $\RGAIC(Q_1,Q_2,\Psi)$ denote the  sum rate achievable by treating interference as noise for the genie-aided interference channel using Gaussian inputs with input covariance matrix $Q_i$ at transmitter $i$:
\[
\RGAIC(Q_1,Q_2,\Psi) = \sum_{i=1}^2 I(\ux_{iG};\uy_{iG},\us_{iG}).
\]
where $\Psi$ is used to denote the genie parameters. Let $\CsumGAIC(\Psi)$ denote the sum capacity of the genie-aided interference channel. The steps required in the outer bound are as follows:
\begin{equation} \label{eq:proof}
\begin{split}
\CsumIC \stackrel{(a)} \leq & \ \CsumGAIC(\Psi) \\
\stackrel{(b)}		  = & \ \max_{Q_{i} \in \mathcal{Q}_{i}} \RGAIC(Q_1,Q_2,\Psi) \\
\stackrel{(c)}		  = & \ \RGAIC(Q_1^*,Q_2^*,\Psi) \\
\stackrel{(d)}		  = & \ \RIC(Q_1^*,Q_2^*) \\
							  \stackrel{(e)} \leq & \ \CsumIC.
\end{split}
\end{equation}
Step (a) is obvious since providing the receivers with extra information can only improve the sum capacity. Step (e) follows from Lemma~\ref{eq:max-ric}. Steps (b), (c), and (d) are the non-trivial steps:
\begin{itemize}
\item \textbf{Useful genie:} The genie is chosen to satisfy step (b), i.e., treating interference as noise, and using Gaussian inputs, achieve the sum capacity of the genie-aided interference channel. We say that such a genie is {\em useful}.
\item \textbf{Step (c):} The genie is chosen so that the optimal covariance matrices $(Q_{1}^{*},Q_{2}^{*})$ maximizing the achievable sum-rate of the original interference channel \eqref{eq:optQ-mimo}, also maximize the achievable sum-rate of the genie-aided interference channel.
\item \textbf{Smart genie:} The genie is chosen to satisfy step (d), i.e., it does not give away any extra information when the coding strategy is fixed to using Gaussian inputs with covariance matrices $(Q_{1}^{*},Q_{2}^{*})$ and treating interference as noise. We say that a genie is {\em smart} with respect to (w.r.t.) the covariance matrices $(Q_{1},Q_{2})$ if $\RGAIC(Q_1,Q_2,\Psi) = \RIC(Q_1,Q_2)$. So, step (d) says that the genie is chosen to be smart w.r.t. $(Q_{1}^{*},Q_{2}^{*})$.
\end{itemize}
\medskip
These steps were first introduced in \cite{Shang-Kramer-Chen-IT-2008,Motahari-Khandani-IT-2008,Annapureddy-Veeravalli-IT-2008} for the SISO interference channel, where a genie is constructed satisfying all the steps in \eqref{eq:proof} when the channel parameters satisfy the following condition for low interference regime
\[
\left|\frac{h_{12}}{h_{22}}\left(1 + h_{21}^{2}P_{1}\right)\right| + \left|\frac{h_{21}}{h_{11}}\left(1 + h_{12}^{2}P_{2}\right)\right| \leq 1
\]
In this paper, we intend to generalize this outer bound to MIMO interference channels. The key step, new to the MIMO interference channels, is the optimization step (c), which is implicitly taken care of in the case of SISO interference channel. We now summarize the known results on this problem \cite{shang_capacity_2009,Shang-et-al-Allerton2008}:

\begin{itemize}

\item \textbf{Covariance constraint:}
In \cite{shang_capacity_2009}, Shang et. al. studied the MIMO interference channel with the average transmit covariance constraint 
\begin{equation} \label{eq:cov-constraint}
\mathcal{Q}_{i} = \{Q_{i}: Q_{i} \preceq \tilde{Q}_{i}\}
\end{equation}
instead of the average power constraint give in \eqref{eq:trace-constraint}. With the outer bound techniques used in \cite{Shang-Kramer-Chen-IT-2008,Motahari-Khandani-IT-2008,Annapureddy-Veeravalli-IT-2008,shang_capacity_2009,Shang-et-al-Allerton2008} and this paper, when the genie is useful, the sum-rate $\RGAIC(Q_1,Q_2,\Psi)$ is concave in $(Q_{1},Q_{2})$, and satisfies the following property:
\begin{equation} \label{eq:inc}
\RGAIC(Q_1,Q_2,\Psi) \leq \RGAIC(\tilde{Q}_1,\tilde{Q}_2,\Psi), \forall Q_{1} \preceq \tilde{Q}_{1}, Q_{2} \preceq \tilde{Q}_{2}.
\end{equation}
With covariance constraint \eqref{eq:cov-constraint}, this property \eqref{eq:inc} automatically takes care of the optimization step (c), and thus the problem simplifies to finding conditions under which a genie exists that is useful and smart w.r.t $(\tilde{Q}_{1},\tilde{Q}_{2})$ (See Theorem~6 in \cite{shang_capacity_2009}).

\item \textbf{Power constraint:} In \cite{Shang-et-al-Allerton2008}, Shang et. al. applied the above insight from the covariance constraint problem to the power constraint problem leading to the following observation (See Thereom~1 of \cite{Shang-et-al-Allerton2008}).   If for every $(\tilde{Q}_{1},\tilde{Q}_{2})$ satisfying the power constraint $\tr{\tilde{Q}_{i}} \leq P_{i}$, there exists a genie that is both useful and smart w.r.t. $(\tilde{Q}_{1},\tilde{Q}_{2})$, then using Gaussian inputs and treating interference as noise achieves the sum capacity. 
\end{itemize}
However, the proof outline in \eqref{eq:proof} only requires the existence of a genie that is useful and smart w.r.t. $(Q_{1}^{*},Q_{2}^{*})$. Asking for the existence of such a genie for every $(\tilde{Q}_{1},\tilde{Q}_{2})$ satisfying $\tr{\tilde{Q}_{i}} \leq P_{i}$ can be too restrictive. For example, in the case of MISO interference channel, a genie can be smart w.r.t. $(\tilde{Q}_{1},\tilde{Q}_{2})$ only if they are unit rank. Hence, the condition in \cite{Shang-et-al-Allerton2008} cannot be satisfied by a MISO interference channel no matter how the channel parameters are chosen.

In this paper, we make the following contributions:
\begin{itemize}
\item \textbf{MIMO interference channel:} In Section~\ref{sec:mimo}, we show that if the optimal covariance matrices $(Q_{1}^{*},Q_{2}^{*})$ are full rank, then it is just sufficient to check for the existence of a genie that is useful and smart w.r.t. $(Q_{1}^{*},Q_{2}^{*})$. To better understand why this is so, consider the optimization problem
\[
\max_{a \leq x \leq b} f(x).
\]
Suppose $f(x)$ is not concave, and has a local maximum point at $x^{*} \in (a,b)$. This implies $f'(x^{*}) = 0$. Now, if there exists a concave function $g(x)$ such that $g(x^{*}) = f(x^{*})$, $g(x) \geq f(x), \forall x \in [a,b]$,  then we can show that $x^{*}$ achieves global maximum for both $f(.)$ and $g(.)$. First, observe that $x^{*} \in (a,b)$, and 
\begin{equation} \label{eq:ex1}
x^{*} = \arg\min_{a \leq x \leq b} (g - f)(x)
\end{equation}
which implies that $(g-f)'(x^{*}) = 0$. This, combined with $f'(x^{*}) = 0$, implies that $g'(x^{*}) = 0$. Since $g(.)$ is concave, it follows that $x^{*}$ achieves global maximum of $g(.)$, and 
\[
f(x) \leq g(x) \leq g(x^{*}) = f(x^{*}), \forall a \leq x \leq b.
\]
By replacing $f$ and $g$ with $\RIC$ and $\RGAIC$ respectively, we use the same idea to show that the step (c) in \eqref{eq:proof} is automatically satisfied if the steps (b) and (d) are satisfied and $(Q_{1}^{*},Q_{2}^{*})$ are full rank. To understand why the full rank condition is necessary, observe that the argument in the above example fails if $x^{*} = a$ or $b$, because then \eqref{eq:ex1} does not necessary imply that $g'(x^{*}) - f'(x^{*}) = 0$.

\item \textbf{Symmetric MISO interference channel:} For the MISO interference channel, the optimal covariance matrices $(Q_1^*,Q_2^*)$ are unit rank. So, just the existence of a genie that is useful and smart w.r.t. $(Q_1^*,Q_2^*)$ does not guarantee the validity of the optimization step (c). In Section~\ref{sec:miso}, we consider the symmetric MISO interference channel and show that if the channel parameters satisfy an additional condition, beyond what is required for the existence of useful and smart genie, then step (c) holds. We also derive an explicit threshold condition for the low interference regime.

\item \textbf{Symmetric SIMO interference channel:} In the case of the SIMO interference channel, we have only one transmit antenna, and the optimization step (c) does not even come into the play. Similar to the SISO counterpart, the problem is to only find the conditions under which a genie exists that is both useful and smart. In Section~\ref{sec:simo}, we simplify these conditions in the symmetric case, and show that they result in a threshold condition that is identical to that obtained in the case of the dual MISO interference channel.
\end{itemize}

\section{MIMO Interference Channel} \label{sec:mimo}
Let $\us_{i}$ denote the side information given to the receiver $i$:
\begin{equation}
\begin{split}
\us_1 = H_{21}\ux_1 + \uw_1 \\
\us_2 = H_{12}\ux_2 + \uw_2
\end{split}
\end{equation}
where
\[
\left[ \begin{array}{c} \uz_{i} \\ \uw_{i} \end{array} \right] \sim \mathcal{N}\left(\underline{0}, \left[ \begin{array}{c c} I & A_{i} \\ A_{i}^{\top} & \SigmaGenie_{i} \end{array}\right] \right).
\]
Since the covariance matrix has to be positive semi-definite, the genie parameters have to satisfy
\begin{equation} \label{eq:genie-cond1}
\left[ \begin{array}{c c} I & A_{i} \\ A_{i}^{\top} & \SigmaGenie_{i} \end{array}\right] \succeq 0.
\end{equation}
\begin{lemma}[Theorem 7.7.6 in \cite{horn-jhonson}] \label{lemma:schur}
If $\SigmaGenie \succ 0$, then the following two statements are equivalent:
\[
\left[ \begin{array}{c c} I & A \\ A^{\top} & \SigmaGenie \end{array}\right] \succeq 0
\]
\[
I - A\SigmaGenie^{-1}A^\top \succeq 0.
\]
\end{lemma}
\medskip
Recall that $\Psi$ denotes the genie parameters collectively, and $\RGAIC(Q_1,Q_2,\Psi)$ denotes the sum rate achievable by treating interference as noise for the genie-aided interference channel using Gaussian inputs with input covariance matrix $Q_i$ at transmitter $i$. We now extend Lemmas~12 and 13 in \cite{Annapureddy-Veeravalli-IT-2008} (the so called {\em usefulness} and {\em smartness} conditions) to the MIMO interference channel. Lemmas~\ref{lemma:mimo-useful} and \ref{lemma:mimo-smart} present conditions under which steps (b) and (d) in \eqref{eq:proof} hold true respectively.
\begin{lemma}[Useful Genie]  \label{lemma:mimo-useful}
If the genie ($\Psi$) satisfies the following conditions
\begin{equation} \label{eq:condition-useful}
\begin{split}
\SigmaGenie_1 \preceq & \ I - A_2\SigmaGenie_2^{-1}A^{\top}_2 \\
\SigmaGenie_2 \preceq & \ I - A_1\SigmaGenie_1^{-1}A^{\top}_1
\end{split}
\end{equation}
then the sum capacity of the genie-aided channel is given by
\[
\CsumGAIC(\Psi) =  \max_{Q_{i} \in \mathcal{Q}_{i}}\RGAIC(Q_1,Q_2,\Psi)
\]
and $\RGAIC(Q_1,Q_2,\Psi)$ is concave in $(Q_1,Q_2)$. Furthermore, if the inequalities in \eqref{eq:condition-useful} are met with equality, then
\[
\CsumGAIC(\Psi) =  \max_{Q_{i} \in \mathcal{Q}_{i}}\sum_{i = 1}^2 \entropy{\uy_{iG}|\us_{iG}} - \entropy{\uw_i}.
\]\end{lemma}
\medskip
\begin{proof}
See Appendix~\ref{app:mimo-useful}.
\end{proof}
\begin{remark} \label{remark:useful}
When the cross channel matrices $(H_{12},H_{21})$ are not full-column rank, the conditions \eqref{eq:condition-useful} can be relaxed as we do later in Lemma~\ref{lemma:simo-useful} for the SIMO interference channel.
\end{remark}
\begin{lemma}[Smart Genie]\label{lemma:mimo-smart}
For any input covariance matrices $(Q_1,Q_2)$,
\[
\RGAIC(Q_1,Q_2,\Psi) = \RIC(Q_1,Q_2).
\]
if and only if (iff)
\begin{equation} \label{eq:condition-smart}
\begin{split}
\left(A_1^{\top}(H_{12}Q_2H_{12}^{\top} + I)^{-1}H_{11} - H_{21}\right)Q_1 = & \ 0 \\
\left(A_2^{\top}(H_{21}Q_1H_{21}^{\top} + I)^{-1}H_{22} - H_{12}\right)Q_2 = & \ 0.
\end{split}
\end{equation}
\end{lemma}
\medskip
\begin{proof}
See Appendix~\ref{app:mimo-smart}.
\end{proof}
\begin{theorem} \label{thm:mimo}
The sum capacity of the MIMO interference channel \eqref{eq:mimo-ic} is achieved by using Gaussian inputs and treating interference as noise at the receivers,  and is given by
\[
\Csum = \RIC(Q_1^*,Q_2^*)
\]
if there exists a full rank local optimal solution $(Q_1^*,Q_2^*)$ to the optimization problem
\begin{equation} \label{eq:argmax-ic}
\max_{Q_{i} \in \mathcal{Q}_{i}} \RIC(Q_1,Q_2),
\end{equation}
and  there exist matrices $A_1,A_2,\SigmaGenie_1 \succ 0 ,\SigmaGenie_2 \succ 0$ satisfying the usefulness and smartness conditions corresponding to $(Q_1^*,Q_2^*)$, i.e.,
\begin{equation} \label{eq:conds-thm1}
\begin{split}
\SigmaGenie_1 \preceq I - A_2\SigmaGenie_2^{-1}A^{\top}_2 ~~~~~~~~~ & \ \\
\SigmaGenie_2 \preceq I - A_1\SigmaGenie_1^{-1}A^{\top}_1 ~~~~~~~~~ & \ \\
\left(A_1^{\top}(H_{12}Q_2^*H_{12}^{\top} + I)^{-1}H_{11} - H_{21}\right)Q_1^* = & \ 0 \\
\left(A_2^{\top}(H_{21}Q_1^*H_{21}^{\top} + I)^{-1}H_{22} - H_{12}\right)Q_2^* = & \ 0.
\end{split}
\end{equation}
\end{theorem}
\medskip
\begin{proof}
Since
\[
\begin{split}
I - A_2\SigmaGenie_2^{-1}A^{\top}_2 \succeq \SigmaGenie_1 \succ 0 \\
I - A_1\SigmaGenie_1^{-1}A^{\top}_1 \succeq \SigmaGenie_2 \succ 0
\end{split}
\]
from Lemma~\ref{lemma:schur},  condition \eqref{eq:genie-cond1} is satisfied. The outline of the proof is given in \eqref{eq:proof}. Here, we justfy the steps (b), (c) and (d) in \eqref{eq:proof}. Steps (b) and (d) follow from Lemmas~\ref{lemma:mimo-useful} and \ref{lemma:mimo-smart} respectively, and it only remains to prove step (c). The Lagrangian associated with \eqref{eq:argmax-ic} is \cite{boyd-convex-optimization}:
\[
\begin{split}
-\RIC(Q_1,Q_2) + \sum_{i=1}^2\lambda_i(\tr{Q_i} - P_i) - \tr{M_iQ_i}.
\end{split}
\]
Since $(Q_1^*,Q_2^*)$ is a local optimal solution to \eqref{eq:argmax-ic}, there exist dual variables $\lambda_i \geq 0$ and $M_i \succeq 0$ satisfying the KKT conditions \cite{boyd-convex-optimization}:
\begin{equation} \label{eq:kkt1}
\begin{split}
\nabla_{Q_i}\RIC(Q_1^*,Q_2^*) = & \  \lambda_iI - M_i, \\
\lambda_i(\tr{Q_i^*} - P_i) = & \ 0,\\
\tr{M_iQ_i^*} = & \ 0.
\end{split}
\end{equation}
Now, let $\RDIFF(Q_1,Q_2,\Psi)$ denote the extra sum rate achievable because of the genie:
\[
\begin{split}
\RDIFF(Q_1,Q_2,\Psi) 	= & \ \RGAIC(Q_1,Q_2,\Psi) - \RIC(Q_1,Q_2) \\
											= & \ \sum_{i=1}^{2} I(\ux_{iG};\uy_{iG}|\us_{iG}) \\
										 \geq & \ 0.
\end{split}
\]
Since the genie $\Psi$ satisfies condition \eqref{eq:condition-smart}, from Lemma~\ref{lemma:mimo-smart} we have
\[
\RDIFF(Q_1^*,Q_2^*,\Psi) = 0.
\]
Therefore $(Q_1^*,Q_2^*)$ is also a solution to the optimization problem
\begin{equation} \label{eq:argmin-diff}
\min_{Q_i \succeq 0} \RDIFF(Q_1,Q_2,\Psi).
\end{equation}
The Lagrangian associated with \eqref{eq:argmin-diff} is
\[
\begin{split}
\RDIFF(Q_1,Q_2,\Psi) - \sum_{i=1}^2 \tr{M_iQ_i}.
\end{split}
\]
There exists dual variables $N_i \succeq 0$ satisfying the KKT conditions
\[
\begin{split}
\nabla_{Q_i}\RDIFF(Q_1^*,Q_2^*) = & \ N_i, \\
\tr{N_iQ_i^*} = & \ 0.
\end{split}
\]
Since $Q_i^*$ is full rank,  $\tr{N_iQ_i^*} = 0$ implies $N_i = 0$, and hence
\begin{equation} \label{eq:kkt2}
\nabla_{Q_i}\RDIFF(Q_1^*,Q_2^*) = 0
\end{equation}
Combining  \eqref{eq:kkt1} and \eqref{eq:kkt2}, we have
\begin{equation} \label{eq:kkt-gaic}
\begin{split}
\nabla_{Q_i}\RGAIC(Q_1^*,Q_2^*) = & \lambda_iI - M_{i}, \\
\lambda_i(\tr{Q_i^*} - P_i) = & \ 0, \\
\tr{M_iQ_i^*} = & \ 0
\end{split}
\end{equation}
which are nothing but the KKT conditions associated with the problem
\begin{equation} \label{eq:argmax-gaic}
\max_{Q_{i} \in \mathcal{Q}_{i}} \RGAIC(Q_1,Q_2,\Psi).
\end{equation}
From Lemma~\ref{lemma:mimo-useful}, $\RGAIC(Q_1,Q_2)$ is concave in $Q_1$ and $Q_2$. Since the objective function is concave, the KKT conditions are not only necessary but also sufficient, and hence $(Q_1^*,Q_2^*)$ is a globally optimal solution to the optimization problem \eqref{eq:argmax-gaic}. (See Section 5.5.3 in \cite{boyd-convex-optimization} for an explanation). This verifies step (c) of \eqref{eq:proof}, and completes the proof of Theorem~\ref{thm:mimo}. 
\end{proof}
\section{MISO Interference Channel}
\label{sec:miso}
Consider a symmetric two-user Gaussian MISO interference channel:
\begin{equation} \label{eq:miso-ic}
\begin{split}
Y_{1} = \ud^{\top}\ux_{1} + h\uc^{\top}\ux_{2} + Z_{1} \\
Y_{2} = \ud^{\top}\ux_{2} + h\uc^{\top}\ux_{1} + Z_{2}
\end{split}
\end{equation}
obtained by making the following substitutions in \eqref{eq:mimo-ic}:
\[
\begin{split}
H_{11} = H_{22} = & \ \ud^{\top} \\
H_{12} = H_{21} = & \ h\uc^{\top} \\
P_{1} = P_{2} = & \ P
\end{split}
\]
where $\ud$ and $\uc$ are unit norm vectors, denoting the directions of direct link and cross link respectively. Without any loss of generality, we can assume two transmit antennas and
\[
\begin{split}
\uc = & \ [1 \ 0]^\top \\
\ud =  & \ [\ct \ \st]^\top, \ \theta \in (0,\pi/2).
\end{split}
\]
by projecting $\ux_1$ and $\ux_2$ along appropriate basis vectors. See Appendix~\ref{app:twoAnts} for an explanation. The extreme case $\theta = 0$ corresponds to the SISO interference channel, which was considered in \cite{Annapureddy-Veeravalli-IT-2008}. The extreme case $\theta = \pi/2$ corresponds to the no interference scenario, with the capacity region given by $\{(R_{1},R_{2}):R_{i} \leq \frac{1}{2}\logfn{1 + P}\}$.
\subsection{Achievable sum-rate} \label{sec:misoa}
The sum-rate achieved by  using Gaussian inputs with input covariance matrix $Q_i$ at transmitter $i$ and treating interference as noise at the receivers is
\[
\begin{split}
\RIC(Q_1,Q_2) = & \ \sum_{i=1}^2 I(\ux_{iG};\uy_{iG}) \\
= & \ \frac{1}{2}\logfn{1 + \frac{\ud^{\top}Q_{1}\ud}{1 + h^{2}\uc^{\top}Q_{2}\uc}} + \frac{1}{2}\logfn{1 + \frac{\ud^{\top}Q_{2}\ud}{1 + h^{2}\uc^{\top}Q_{1}\uc}}
\end{split}
\]
This problem of finding the optimal transmit beams for the MISO interference channel is studied in \cite{shang_multi-user_2009,jorswieck_complete_2008} and \cite{zakhour}. In \cite{shang_multi-user_2009}, it is shown that the optimal covariance matrices $Q_{1}^{*}$ and $Q_{2}^{*}$ are unit-rank. Even though the channel is symmetric across the users, since the sum-rate $\RIC(Q_1,Q_2)$ is not necessarily concave in $(Q_{1},Q_{2})$, the optimal covariance matrices $Q_{1}^{*}$ and $Q_{2}^{*}$ need not be the same. In this Section, for the purpose of achievable sum-rate, we restrict ourselves to covariance matrices symmetric across the users, i.e, 
\[
Q_{1} = Q_{2} = Q = P\ub\ub^{\top}
\] 
where $\ub$ is the unit norm transmit beamforming vector. With this assumption, the achievable sum-rate is given by
\[
\RIC(Q_1,Q_2) = \logfn{1 + \textrm{SINR}}
\]
where SINR denotes the signal to interference and noise ratio:
\[
\textrm{SINR} = \frac{P(\ubd)^2}{1 + h^2P(\ubc)^2} = \frac{\ub^{\top}(P\ud\ud^{\top})\ub}{\ub^{\top}(I + h^2P\uc\uc^{\top})\ub}.
\]
The optimal $\ub$ maximizing SINR is given by the generalized eigenvector of the matrix pair 
\[
\left(P\ud\ud^{\top},\textrm{I} + h^{2}P\uc\uc^{\top} \right)
\]
corresponding to the largest generalized eigenvalue, given by
\begin{equation} \label{eq:optBeam}
\ub = \frac{\inv{I + h^2P\uc\uc^{\top}}\ud}{||\inv{I + h^2P\uc\uc^{\top}}\ud||}.
\end{equation}
The corresponding SINR is given by
\begin{equation} \label{eq:sinr}
\begin{split}
\textrm{SINR} = & \ \frac{P(\ubd)^2}{\ub^{\top}\left(I + h^2P\uc\uc^{\top}\right)\ub}\\
					 = & \ P\ud^{\top}\inv{I + h^2P\uc\uc^{\top}}\ud \\
\stackrel{(a)} = & \ P\ud^{\top}\left(I - \frac{h^2P}{1+h^2P}\uc\uc^{\top}\right)\ud \\
					 = & \ P\left(1 - \frac{h^2P\cst}{1 + h^2P}\right) \\
					 = & \ \frac{P\cst}{1 + h^2P} + P\sst.
\end{split}
\end{equation}
where step (a) follows from the matrix inversion lemma:
\begin{equation} \label{eq:matrix-inversion-lemma}
\inv{A + UCV} = A^{-1} - A^{-1}U\inv{C^{-1} + VA^{-1}U}VA^{-1}.
\end{equation}
Therefore, a sum-rate of
\[
\RIC(Q^*,Q^*)  = \logfn{1 + \textrm{SINR}} = \logfn{1 + \frac{P\cst}{1 + h^2P} + P\sst }
\]
is achievable. 

We now characterize the low interference regime for the MISO interference channel,  and provide explicit conditions on $h$ and $\theta$, when $\RIC(Q^*,Q^*)$ is equal to the sum capacity. We cannot use Theorem~\ref{thm:mimo} to do this, because the optimal covariance matrices $Q_{1}^{*}$ and $Q_{2}^{*}$ are not full-rank, as required in the hypothesis of Theroem~\ref{thm:mimo}. 
\subsection{Low Interference Regime}
\begin{theorem} \label{thm:miso-lowInterferenceRegime}
The sum capacity of the MISO interference channel \eqref{eq:miso-ic} is achieved by using Gaussian inputs, transmit beamforming along the vector $\ub$ given in \eqref{eq:optBeam}, and treating interference as noise at the receivers, and is given by
\[
\CsumIC = \logfn{1 + \frac{P\cst}{1 + h^2P} + P\sst }
\]
if the channel gain parameter $h$ satisfies:
\begin{equation}\label{eq:miso_threshold}
|h| < h_0(\theta,P)
\end{equation}
with  $h_0(\theta,P)$ being the positive solution to the implicit equation
\begin{equation} \label{eq:condition-simo}
h^2 - \sst = \left(\frac{\ct}{1 + h^2P} - h \right)_+^2.
\end{equation}
(The notation $x_+^2$ is used to denote $(\max(0,x))^2$.)
\end{theorem}
\begin{proof}
First, observe that the capacity region of the symmetric MISO interference channel \eqref{eq:miso-ic} does not depend on the sign of the channel parameter $h$.  By 
replacing $\ux_{2}$ and $Y_{2}$ with $-\ux_{2}$ and $-Y_{2}$ respectively, we can convert the interference channel with negative $h$ to the interference channel with positive $h$. Therefore, with out any loss of generality, we assume that $h \geq 0$.

The achievability part of Theorem~\ref{thm:miso-lowInterferenceRegime} is  established in Section~\ref{sec:misoa}:
\[
\CsumIC \geq \RIC(Q^*,Q^*) = \logfn{1 + \frac{P\cst}{1 + h^2P} + P\sst }.
\]
We now prove the converse, i.e., that
\[
\CsumIC \leq \RIC(Q^*,Q^*).
\]
The outline of the converse proof is given in \eqref{eq:proof}. We justify the steps (b), (c) and (d) of \eqref{eq:proof} here. Specializing the genie \eqref{eq:genie-mimo} to the MISO interference channel, we get
\begin{equation} \label{eq:genie-miso}
\begin{split}
S_{1} = & \ h\uc^{\top}\ux_{1} + W_{1} \\
S_{2} = & \ h\uc^{\top}\ux_{2} + W_{2}.
\end{split}
\end{equation}
We restrict the genie to be symmetric across the users, with the genie parameters $\Psi = \{\SigmaGenie,a\}$ chosen to satisfy the usefulness condition \eqref{eq:condition-useful} of Lemma~\ref{lemma:mimo-useful} and the smartness condition \eqref{eq:condition-smart} of Lemma~\ref{lemma:mimo-smart}. From \eqref{eq:condition-smart}, we get
\[
\begin{split}
\left(\frac{a}{1 + h^{2}P|\optbf^{\top}\uc|^{2}}\ud^{\top} - h\uc^{\top}\right)Q^{*} = & \ \ 0 \\
\left(\frac{a\ud^{\top}}{1 + h^{2}P|\optbf^{\top}\uc|^{2}} - h\uc^{\top}\right)\optbf = & \  0
\end{split}
\]
and thus
\[
a =\frac{h\uc^{\top}\ub}{\ud^{\top}\ub}(1 + h^{2}P|\optbf^{\top}\uc|^{2}).
\]
From \eqref{eq:condition-useful}, we get
\[
\begin{split}
\SigmaGenie \leq 1 - \frac{a^2}{\SigmaGenie} \\
\SigmaGenie^2 - \SigmaGenie + a^2 \leq 0
\end{split}
\]
We choose
\[
\SigmaGenie = 0.5 + 0.5\sqrt{1 - 4a^2}
\] so that \eqref{eq:condition-useful} is satisfied with equality. Thus
\begin{equation} \label{eq:psi-miso}
\begin{split}
a = & \ \expect{W_iZ_i} = \ h\frac{\ub^{\top}\uc}{\ub^{\top}\ud}\left(1 + h^{2}P|\optbf^{\top}\uc|^{2}\right) \\
\SigmaGenie = & \ \expect{W_i^2} = 0.5 + 0.5\sqrt{1 - 4a^2}.
\end{split}
\end{equation}
In Appendix~\ref{sec:eqv-threshold}, we show that $h < h_{0}(\theta,P)$ implies that 
\begin{equation}
\label{eq:thrcond1}
a < 0.5
\end{equation}
Thus, a real $\SigmaGenie > 0$ exists, and the conditions of Lemmas~\ref{lemma:mimo-useful} and \ref{lemma:mimo-smart} are met. The steps (b) and (c) of \eqref{eq:proof} follow from Lemmas~\ref{lemma:mimo-useful} and \ref{lemma:mimo-smart} respectively. Thus, it only remains to verify that step (c) of \eqref{eq:proof} holds true, i.e.,
\[
(Q^*,Q^*) = \arg\max_{Q_{i} \in \mathcal{Q}_{i}} \RGAIC(Q_1,Q_2,\Psi).
\]
Since the condition \eqref{eq:condition-useful} of Lemma~\ref{lemma:mimo-useful} is met with equality, we have
\[
\RGAIC(Q_1,Q_2,\Psi) = \sum_{i = 1}^2 \entropy{\uy_{iG}|\us_{iG}} - \entropy{W_i}.
\]
Furthermore, $\RGAIC(Q_1,Q_2,\Psi)$ is concave in $(Q_1,Q_2)$. Since the channel and the genie are symmetric across the users $\RGAIC(Q_1,Q_2,\Psi)$ is also symmetric across the users, i.e.,
\[
\RGAIC(Q_1,Q_2,\Psi) = \RGAIC(Q_2,Q_1,\Psi)
\]
Therefore, we have
\[
\begin{split}
\RGAIC(Q_1,Q_2,\Psi) = & \ \frac{1}{2}\RGAIC(Q_1,Q_2,\Psi) + \frac{1}{2}\RGAIC(Q_2,Q_1,\Psi)\\
									\leq & \ \RGAIC(Q,Q,\Psi),
\end{split}
\]
where
\[
Q = \left(\frac{Q_1 + Q_2}{2}\right).
\]
Since the power constraints are also symmetric across users
\[
\mathcal{Q}_1 = \mathcal{Q}_2 = \mathcal{Q} = \{Q: Q \succeq 0, \tr{Q} \leq P\}
\]
$Q_1 \in \mathcal{Q}$ and $Q_2 \in \mathcal{Q}$ implies $Q \in \mathcal{Q}$. Therefore, we have shown that having symmetric $Q$ across the users maximizes $\RGAIC(Q_1,Q_2,\Psi)$. Hence it is sufficient to prove that
\begin{equation} \label{eq:armgax-gaic-miso}
Q^* = \arg\max_{Q \in \mathcal{Q}} \RGAIC(Q,Q,\Psi)
\end{equation}
Now,
\begin{equation} \label{eq:step1}
\begin{split}
\RGAIC(Q,Q,\Psi) = & \ \sum_{i = 1}^2 \entropy{\uy_{iG}|\us_{iG}} - \entropy{W_i} \\
= & \ \sum_{i = 1}^2 \entropy{\uy_{iG} - \mu\us_{iG}|\us_{iG}} - \entropy{W_i} \\
\stackrel{(a)}\leq & \ \sum_{i = 1}^2 \entropy{\uy_{iG} - \mu\us_{iG}} - \entropy{W_i} \\
= & \ \log\left(\tr{MQ} \ + k^2\right) - \log(\SigmaGenie) \\
\end{split}
\end{equation}
where step (a) follows from the fact that conditioning reduces entropy, and the matrix $M$ and the constant $c$ are given by
\begin{equation} \label{eq:M}
\begin{split}
M = & \ (\ud - \mu{}h\uc)(\ud - \mu{}h\uc)^{\top} + h^2\uc\uc^{\top} \\
k^2 = & \ \expect{(Z_1 - \mu{}W_1)^2}.
\end{split}
\end{equation}
The constant $\mu$ is chosen such that the step (a) is tight when $Q = Q^*$, which is true if $\mu{}S_{iG}$ is the MMSE estimate of $Y_{iG}$ given $S_{iG}$ when $Q = Q^*$, i.e., if
\begin{equation} \label{eq:mu}
\begin{split}
\mu = & \ \frac{\expect{S_{iG}Y_{iG}}}{\expect{S_{iG}^2}}\\
 = & \ \frac{a + hP\ub^{\top}\uc\ub^{\top}\ud}{\SigmaGenie + h^2P(\ub^{\top}\uc)^2}.
\end{split}
\end{equation}
In Appendix~\ref{sec:eqv-threshold}, we show that $h < h_{0}(\theta,P)$ implies
\begin{equation}
\label{eq:thrcond2}
\mu{}h < \cos(\theta)
\end{equation} 
In Appendix~\ref{sec:eigenValue}, we show that $\mu{}h < \cos(\theta)$ implies
\begin{equation} \label{eq:step2}
\tr{MQ} \leq \tr{MQ^{*}}, \forall Q \in \mathcal{Q}.
\end{equation}
From \eqref{eq:step1} and \eqref{eq:step2}, we have
\[
\RGAIC(Q,Q,\Psi) \leq \ \log\left(\tr{MQ^*} \ + k^2\right) - \log(\SigmaGenie)
\]
Moreover, since all the steps in \eqref{eq:step1} and \eqref{eq:step2} are tight when $Q = Q^*$, we have
\[
\begin{split}
\RGAIC(Q^*,Q^*,\Psi) = & \ \ \log\left(\tr{MQ^*} \ + k^2\right) - \log(\SigmaGenie) \\
\RGAIC(Q,Q,\Psi) \leq & \ \RGAIC(Q^*,Q^*,\Psi), \forall Q \in \mathcal{Q}.
\end{split}
\]
This verifies the step (c) of \eqref{eq:proof} and completes the converse proof.
\end{proof}
In the following section we establish that results identical to those in Thereom~\ref{thm:miso-lowInterferenceRegime} also hold true for the dual SIMO interference channel, although the proofs are quite different.
\section{SIMO Interference Channel} \label{sec:simo}
Consider the symmetric SIMO interference channel
\begin{equation} \label{eq:simo-ic}
\begin{split}
\uy_{1} = \ud{}X_{1} + h\uc{}X_{2} + \uz_{1} \\
\uy_{2} = \ud{}X_{2} + h\uc{}X_{1} + \uz_{2}
\end{split}
\end{equation}
obtained by making the following substitutions in \eqref{eq:mimo-ic}:
\[
\begin{split}
H_{11} = H_{22} = & \ \ud \\
H_{12} = H_{21} = & \ h\uc \\
P_{1} = P_{2} = & \ P
\end{split}
\]
where $\ud$ and $\uc$ are unit norm vectors, denoting the directions of direct link and cross link respectively. Similar to the MISO Interference channel, without any loss of generality, we assume two receive antennas and
\[
\begin{split}
\uc = & \ [1 \ 0]^\top \\
\ud =  & \ [\ct \ \st]^\top, \ \theta \in (0,\pi/2).
\end{split}
\]
by projecting $\uy_1$ and $\uy_2$ along appropriate basis vectors. See Appendix~\ref{app:twoAnts} for an explanation.
\subsection{Achievable sum-rate}
\label{sec:simoa}
Assume that each receiver does performs beamforming along the vector $\ub$ given in \eqref{eq:optBeam}. By treating interference as noise at the receivers, the following signal to interference and noise ratio (SINR) is achieved.
\[
\begin{split}
\textrm{SINR} = & \ \frac{P(\ubd)^2}{1 + h^2P(\ubc)^2} \\
					 		= & \ \frac{P\cst}{1 + h^2P} + P\sst.
\end{split}
\]
Therefore, a sum-rate of
\[
\RIC(P,P)  = \logfn{1 + \textrm{SINR}} = \logfn{1 + \frac{P\cst}{1 + h^2P} + P\sst }
\]
is achievable.
\subsection{Low Interference Regime}
\begin{theorem} \label{thm:simo-lowInterferenceRegime}
The sum capacity of the SIMO interference channel \eqref{eq:simo-ic} is achieved by using Gaussian inputs, receive beamforming along the vector $\ub$ given in \eqref{eq:optBeam}, and treating interference as noise at the receiver, and is given by
\[
\CsumIC = \logfn{1 + \frac{P\cst}{1 + h^2P} + P\sst }
\]
if $|h| < h_0(\theta,P)$, where $h_0(\theta,P)$ is the positive solution to the implicit equation
\begin{equation} \label{eq:condition-miso}
h^2 - \sst = \left(\frac{\ct}{1 + h^2P} - h \right)_+^2.
\end{equation}
where the notation $x_+^2$ is used to denote $(\max(0,x))^2$.
\end{theorem}
\begin{proof}
First, observe that the capacity region of the symmetric SIMO interference channel \eqref{eq:simo-ic} does not depend on the sign of the channel parameter $h$.  By 
replacing $X_{2}$ and $\uy_{2}$ with $-X_{2}$ and $-\uy_{2}$ respectively, we can convert the interference channel with negative $h$ to the interference channel with positive $h$. Therefore, with out any loss of generality, we assume that $h \geq 0$.

The achievability part of Theorem~\ref{thm:simo-lowInterferenceRegime} is established in Section~\ref{sec:simoa}, i.e.,
\[
\CsumIC \geq \RIC(P,P) = \logfn{1 + \frac{P\cst}{1 + h^2P} + P\sst }.
\]
We now prove the converse, i.e., that
\[
\CsumIC \leq \RIC(P,P).
\]
The outline of the converse proof is given in \eqref{eq:proof}. Specializing the genie \eqref{eq:genie-mimo} to the SIMO interference channel, we get
\begin{equation} \label{eq:genie-simo-def}
\begin{split}
\us_{1} = & \ h\uc{}X_{1} + \uw_{1} \\
\us_{2} = & \ h\uc{}X_{2} + \uw_{2}.
\end{split}
\end{equation}
We restrict the genie to be symmetric across the users, with the genie parameters $\Psi = \{\SigmaGenie,A\}$ defined as be chosen to be
\begin{equation}
\label{eq:genie-simo}
\begin{split}
A = & \ \expect{\uz_i\uw_i^\top} = \ k\vv\uc^\top \\
\SigmaGenie = & \ \expect{\uw_i\uw_i^\top} = \ \eta{}I
\end{split}
\end{equation}
where $k \geq 0$ and $\vv$ is unit norm vector. For the genie parameters to be valid, they have to satisfy \eqref{eq:genie-cond1}, and from Lemma~\ref{lemma:schur}, \eqref{eq:genie-cond1} is true iff
\begin{equation} \label{eq:genie-cond2}
\eta \geq k^2
\end{equation}
As mentioned in Remark~\ref{remark:useful}, Lemma~\ref{lemma:mimo-useful} can be improved for the SIMO interference channel since the cross channel matrices are not full column rank.
\begin{lemma}[Useful Genie for SIMO Channel] \label{lemma:simo-useful}
If the genie ($\Psi$) satisfy
\begin{equation} \label{eq:simo-useful}
\inv{\uc^{\top}\SigmaGenie^{-1}\uc} \leq \inv{\uc^{\top}\inv{I - A\SigmaGenie^{-1}A^{\top}}\uc}
\end{equation}
then the sum capacity of the genie-aided channel is given by
\[
\CsumGAIC(\Psi) =  \RGAIC(P,P,\Psi).
\]
\end{lemma}
\medskip
\begin{proof}
See Appendix~\ref{app:simo-useful}.
\end{proof}
Substituting \eqref{eq:genie-simo} in \eqref{eq:simo-useful}, we get
\begin{equation} \label{eq:simo-useful1}
\eta \leq \frac{\eta - k^2}{\eta - k^2(1 - (\ucv)^2)}.
\end{equation}
From Lemma~\ref{lemma:mimo-smart}, if the condition \eqref{eq:condition-smart} is satisfied at $Q_1 = Q_2 = P$, i.e., if
\[
A^{\top}\inv{h^2P\uc\uc^{\top} + I}\ud - h\uc = 0
\]
equivalently, if
\begin{equation} \label{eq:simo-smart}
h = k\vv^\top\inv{I + h^2P\uc\uc^{\top}}\ud
\end{equation}
then
\[
\RGAIC(P,P,\Psi) = \RIC(P,P).
\]
Therefore, if there exist $\eta,k$ and $\vv$ satisfying \eqref{eq:genie-cond2}, \eqref{eq:simo-useful1} and \eqref{eq:simo-smart}, then we have the required outer bound.
\[
\begin{split}
\CsumIC \leq & \ \CsumGAIC \\
				   = & \ \RGAIC(P,P,\Psi) \\
				   = & \ \RIC(P,P).
\end{split}
\]
The existence of $\eta,k$ and $\vv$ satisfying \eqref{eq:genie-cond2}, \eqref{eq:simo-useful1} and \eqref{eq:simo-smart} is proved in Appendix~\ref{app:simo}, using the threshold condition of the theorem.
\end{proof}
\section{Low Interference Regime}
\label{sec:discussion}
In Sections~\ref{sec:miso} and \ref{sec:simo}, we proved that using Gaussian inputs and treating interference as noise is sum capacity achieving for the symmetric MISO and SIMO interference channels if $h \leq h_0(\theta,P)$ where $h_0(\theta,P)$ is the positive solution to the equation
\begin{equation} \label{eq:condition-miso-simo}
h_0^2 - \sst = \left(\frac{\ct}{1 + h_0^2P} - h_0 \right)_+^2.
\end{equation}
\begin{figure}[t]
\centering
\includegraphics[width=3in]{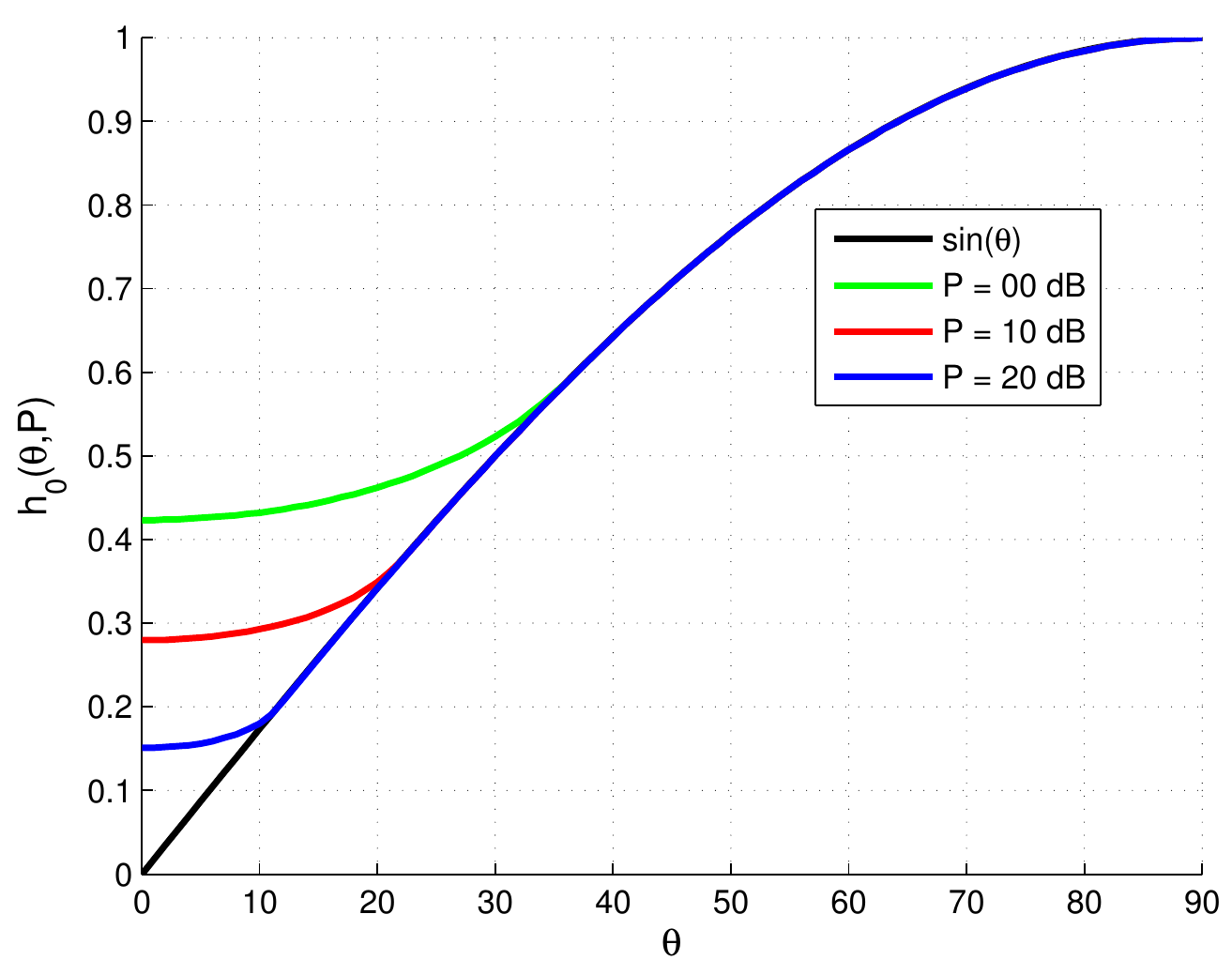}
\caption{Threshold on $h$ charecterizing the low interference regime of MISO and SIMO interference channels}
\label{fig:threshold}
\end{figure}
Observe that, with $\theta = 0$, the threshold $h_{0}(0,P)$ is the solution to the equation
\[
\begin{split}
h_{0}(1 + h_{0}^{2}P) \leq 0.5
\end{split}
\]
which is consistent with the threshold condition for the low interference regime of the SISO interference channel \cite{Annapureddy-Veeravalli-IT-2008}. The threshold $h_0(\theta,P)$ is plotted as a function of $\theta$ for different values of $P$ in Figure~\ref{fig:threshold}. It can be observed that the threshold curve is always above the $\sin(\theta)$ curve and approaches the $\sin(\theta)$ curve as $P$ becomes larger. These observations are summarized in the following Lemma.
\begin{lemma}
The threshold $h_0(\theta,P)$ as defined in \eqref{eq:condition-miso-simo} satisfies the following properties.
\begin{enumerate}
\item Independent of the value of $P$, 
\[
h_0(\theta,P) \geq \st
\]
\item For any fixed $\theta$, 
\[
\lim_{P \rightarrow \infty}h_0(\theta,P) = \st
\]
\end{enumerate}
\end{lemma}
\medskip
\begin{proof}
Any $h_0 < \st$ cannot satisfy \eqref{eq:condition-miso-simo} since the RHS of \eqref{eq:condition-miso-simo} cannot be negative. Therefore, we have the first property. For any $h_0 \geq \st$, 
\[
\frac{\ct}{1 + h_0^2P} - h_0 \leq \frac{\ct}{1 + P\sst{}} - \st
\]
which is less than zero for sufficiently large $P$. Therefore, the RHS of \eqref{eq:condition-miso-simo} is zero for any $h_0 \geq \st$ and hence $h_0$ cannot be greater than $\st$ for sufficiently large $P$. This proves the second property.
\end{proof}
Theorems~\ref{thm:miso-lowInterferenceRegime} and \ref{thm:simo-lowInterferenceRegime} along with the first property of the above Lemma lead to the following corollary. 
\begin{corrly}
The sum capacity of the symmetric MISO/SIMO interference channel is achieved by using Gaussian inputs,  transmit/receive beamforming, and treating interference as noise at the receivers, if $h \leq \st$.
\end{corrly}
\section{Conclusions} \label{sec:concl}
In this paper, we derived sufficient conditions under which using Gaussian inputs and treating interference as noise at the receivers achieves the sum capacity of the two-user MIMO interference channel. The outer bound is based on a genie giving side information to the receivers.  The genie is carefuly chosen to satisfy the following properties.
\begin{itemize}
\item The genie should be useful, i.e., using Gaussian inputs and treating interference as noise at the receivers should achieve the sum capacity of the genie-aided channel.
\item The genie should be smart at $(Q_1^*,Q_2^*)$, which are the optimal covariance matrices that maximize the achievable sum-rate of the interference channel. That is, when Gaussian inputs with covariance matrices $(Q_1^*,Q_2^*)$ are used and interference is treated as noise at the receivers, the presence of the genie should not improve the achievable sum-rate.
\item The optimal covariance matrices $(Q_1^*,Q_2^*)$ for the original interference channel should also maximize the achievable sum-rate of the genie-aided channel.
\end{itemize}
The last property is the new feature of the MIMO interference channel relative to the SISO setting. One of the main contributions of this paper is to show that if $(Q_1^*,Q_2^*)$ are full rank and if there exists a genie satisfying the first two properties, then the last property is automatically satisfied. If $(Q_1^*,Q_2^*)$ are not full rank, then the genie might have to satisfy some additional constraints for the last property to be true. 

The MISO interference channel is an example where the optimal covariance matrices $(Q_1^*,Q_2^*)$ are not full rank, since transmit beamforming maximizes the achievable sum-rate. For the special case of the symmetric MISO interference channel, we derived the additional constraints and derived a threshold condition for the low interference regime. We also showed that the same threshold condition characterizes the low interference regime for the dual SIMO interference channel. The threshold condition is given by $h \leq h_0(\theta,P)$, where $h_0(\theta,P)$ is larger than $\st$. This means that the low interference regime can be quite significant. For example, when $\theta = 45^o$, then using Gaussian inputs and treating interference as noise at  the receivers achieves the sum capacity if INR $( = h^2P)$  is $3$~dB less than SNR $( = P)$. 
\section*{Acknowledgments} 
We  would like to thank Sriram Vishwanath for helping us formulate the MIMO interference channel problem, particularly in symmetric MISO setting. We would also like to thank Ashish Khisti for useful discussions.
\appendix
\subsection{Some useful results}
We now present a few results that will be useful in this paper.
\begin{lemma}[Lemma~1 in \cite{Annapureddy-Veeravalli-IT-2008}]
Let $\ux$ be a random vector, and let $\uy$ and $\us$ be noisy observations of $\ux$.
\[
\begin{split}
\uy & =  \bbA \; \ux + \uz \\
\us & =  \bbC \; \ux + \uw
\end{split}
\]
where $\uz$ and $\uw$ are correlated, zero-mean, Gaussian random vectors, and $\bbA$ and $\bbC$ are real valued matrices. Consider the random vector sequence $\ux^n = (\ux_1, \ldots, \ux_n)$ with the covariance constraint $\frac{1}{n}\sum_{j=1}^{n} \Sigma_{xj} \preceq \Sigma_x$, where $\Sigma_{xj}$ is the covariance matrix of $\ux_j$. Furthermore, let $\uy^n$ and $\us^n$ be the corresponding observations when the noise vector sequences $\uz^n$ and $\uw^n$ each have components that are i.i.d. in time. Then, we have
\[
\entropy{\uy^n|\us^n} \leq n\entropy{\uy_G|\us_G}
\]
where $\uy_G$ and $\us_G$ are $\uy$ and $\us$ when $\ux = \ux_G \sim \mathcal{N}(\uzero,\Sigma_x)$.
\label{lemma1}
\end{lemma}
\begin{lemma}[Lemma~4 in \cite{Annapureddy-Veeravalli-IT-2008}]
Let $\ux^n$ be a random vector sequence with an average covariance constraint, i.e., $\sum_{j=1}^{n}\Sigma_{xj} \preceq n\Sigma_x$, and let $\uz^n$ be an independent random vector sequence, with components that are i.i.d. ${\cal N} (\uzero, \Sigma_z)$. Then
\[
\entropy{\ux^n} - \entropy{\ux^n+\uz^n} \leq n\entropy{\ux_G} - n\entropy{\ux_G+\uz}
\]
where $\ux_G \sim {\cal N} (\uzero,\Sigma_x)$, and equality is achieved if $\ux^n = \ux_G^n$, where $\ux_G^n$ denotes the random sequence with components that are i.i.d. ${\cal N} (\uzero, \Sigma_x)$.
\label{lemma4}
\end{lemma}
\begin{lemma}[Lemma~6 in \cite{Annapureddy-Veeravalli-IT-2008}]
Let $\ux^n$ be a random vector sequence,  and let $\uz^n$ and $\uw^n$ be (possibly correlated) zero-mean Gaussian random vector sequences, independent of $\ux^n$ and i.i.d. in time. Then
\[
\entropy{\ux^n + \uz^n|\uw^n} = \entropy{\ux^n + \uv^n}
\]
where $\uv^n$ is i.i.d. $ \mathcal{N}\left(0,\cov{\uz|\uw}\right)$.
\label{lemma6}
\end{lemma}
\begin{lemma} \label{lemma:concavity}
Let $\ux$ be a random vector with covariance matrix $Q$ and $\uy$ and $\us$ be noisy observations of $\ux$.
\[
\begin{split}
\uy = A\ux + \uz \\
\us = B\ux + \uw
\end{split}
\]
where $\uz$ and $\uw$ are correlated, zero-mean, Gaussian random vectors, and $A$ and $B$ are real-valued matrices. Then, $\entropy{\uy_{G}|\us_{G}}$ is concave in $Q$, where $\uy_G$ and $\us_G$ are $\uy$ and $\us$ when $\ux = \ux_G = \mathcal{N}(\uzero,Q)$.
\end{lemma}
\begin{proof}
Let $T$ be a time sharing random variable taking value that takes values $1$ and $2$ with probabilities $\lambda$ and $1 - \lambda$ respectively. Let $\ux_{1G}$, $\ux_{2G}$ and $\uxt_{G}$ be independent Gaussian random vectors with covariance matrices $Q_1$, $Q_2$ and $\lambda{}Q_1 + (1 - \lambda)Q_2$ respectively and let $\uy_{1G}$, $\uy_{2G}$, $\uyt_{G}$, $\us_{1G}$, $\us_{2G}$ and $\ust_G$ denote the corresponding noisy observations. Now,
\[
\begin{split}
\lambda\entropy{\uy_{1G}|\us_{2G}} + (1-\lambda)\entropy{\uy_{2G}|\us_{2G}} = & \ \entropy{\uy_{TG}|\us_{TG},T}\\
\stackrel{(a)}		\leq & \ \entropy{\uy_{TG}|\us_{TG}} \\
\stackrel{(b)}		\leq & \ \entropy{\uyt_{G}|\ust_{G}}
\end{split}
\]
where step (a) follows because conditioning reduces entropy and step (b) follows because Gaussian distribution maximizes the conditional differential entropy for a given covariance constraint \cite[Lemma~1]{Thomas-IT1987}.
\end{proof}
\begin{lemma} \label{lemma:eig}
Let $\vx_1$ an $\vx_2$ be two vectors such that $\vx_1^{\top}\vx_2 > 0$ and let $\vx$ be an eigenvector of the matrix
\[
M = \vx_1\vx_1^{\top} + \vx_2\vx_2^{\top},
\]
then $\vx$ corresponds to the largest eigenvalue iff $(\vx^{\top}\vx_1)(\vx^{\top}\vx_2) > 0$.
\end{lemma}
\begin{proof}
\[
\begin{split}
M\vx = & \ \lambda\vx \\
\vx_1(\vx_1^{\top}\vx) + \vx_2(\vx_2^{\top}\vx) = & \ \lambda\vx
\end{split}
\]
Therefore, we have
\[
\begin{split}
||\vx_1||^2(\vx_1^{\top}\vx) + (\vx_1^{\top}\vx_2)(\vx_2^{\top}\vx) = & \ \lambda(\vx_1^{\top}\vx) \\
||\vx_2||^2(\vx_2^{\top}\vx) + (\vx_1^{\top}\vx_2)(\vx_1^{\top}\vx) = & \ \lambda(\vx_2^{\top}\vx)
\end{split}
\]
i.e.,
\[
\begin{split}
(\vx_1^{\top}\vx_2)(\vx_2^{\top}\vx) = & \ (\lambda - ||\vx_1||^2)(\vx_1^{\top}\vx) \\
(\vx_1^{\top}\vx_2)(\vx_1^{\top}\vx) = & \ (\lambda - ||\vx_2||^2)(\vx_2^{\top}\vx).
\end{split}
\]
Therefore, we have
\[
(\lambda - ||\vx_1||^2)(\lambda - ||\vx_2||^2) = (\vx_1^{\top}\vx_2)^2
\]
and hence
\[
\lambda = \frac{||\vx_1||^2 + ||\vx_2||^2}{2} \pm \frac{\sqrt{(||\vx_1||^2 - ||\vx_2||^2)^2 + 4(\vx_1^{\top}\vx_2)^2}}{2}.
\]
Now, it can be seen that the larger eigenvalue $\lambda_1$ satisfies
\[
\lambda_1 > \max(||\vx_1||^2,||\vx_2||^2)
\]
and hence the corresponding eigenvector satisfies
\[
(\vx^{\top}\vx_1)(\vx^{\top}\vx_2) > 0
\]
Similarly, it can be seen that the smaller eigenvalue $\lambda_2$ satisfies
\[
\lambda_2 < \min(||\vx_1||^2,||\vx_2||^2)
\]
and hence the corresponding eigenvector satisfies
\[
(\vx^{\top}\vx_1)(\vx^{\top}\vx_2) < 0.
\]
\end{proof}
\subsection{Proof of Lemma~\ref{lemma:mimo-useful}}
\label{app:mimo-useful}
Let $\ux_{iG} \sim \mathcal{N}(\uzero,Q_i)$ and $\uy_G$ and $\us_G$ be $\uy$ and $\us$ when $\ux = \ux_{iG}$, where
\[
Q_i = \frac{1}{n}\sum_{t=1}^{n} \expect{\ux_{it}\ux_{it}^{\top}} \in \mathcal{Q}_i.
\]
Starting with Fano's inequality, we have
\begin{equation} \label{eq:usefulStep1}
\begin{split}
n(\CsumGAIC - \epsilon_n) \leq & \sum_{i = 1}^2 I(\ux_i^n;\uy_i^n,\us_i^n) \\
									 = & \sum_{i = 1}^2 I(\ux_i^n;\us_i^n) + I(\ux_i^n;\uy_i^n|\us_i^n) \\
									 = & \sum_{i = 1}^2 \entropy{\us_i^n} - \entropy{\us_i^n|\ux_i^n} \\ ~~~~~~~~~~~~~~~~~~~~~~~~ & \ + \entropy{\uy_i^n|\us_i^n} - \entropy{\uy_i^n|\us_i^n,\ux_i^n} \\		
\stackrel{(a)}	= & \sum_{i = 1}^2 \entropy{\us_i^n} - n\entropy{\uw_i} \\ ~~~~~~~~~~~~~~~~~~~~~~~~ & \ + \entropy{\uy_i^n|\us_i^n} - \entropy{\uy_i^n|\us_i^n,\ux_i^n} \\		
\stackrel{(b)}	\leq & \sum_{i = 1}^2 \entropy{\us_i^n} - n\entropy{\uw_i}  \\ ~~~~~~~~~~~~~~~~~~~~~~~~ & \ + n\entropy{\uy_{iG}|\us_{iG}} - \entropy{\uy_i^n|\us_i^n,\ux_i^n} \\									 							 
\end{split}
\end{equation}
where step (a) follows from the fact that $\us_i$ given $\ux_i$ is Gaussian noise and is independent of $\ux_i$ and step (b) follows from Lemma~\ref{lemma1}. Now consider
\begin{equation} \label{eq:usefulStep2}
\begin{split}
\entropy{\us_1^n} - & \ \entropy{\uy_2^n|\us_2^n,\ux_2^n} \\
= & \ \entropy{H_{21}\ux_1^n + \uw_1^n } - \entropy{H_{21}\ux_1^n + \uz_2^n|\uw_2^n } \\
\stackrel{(c)} = & \ \entropy{H_{21}\ux_1^n + \uw_1^n } - \entropy{H_{21}\ux_1^n + \uv_1^n} \\
\stackrel{(d)} = & \ \entropy{H_{21}\ux_1^n + \uw_1^n } - \entropy{H_{21}\ux_1^n + \uw_1^n + \uvt_1^n} \\
               = & \ -I(\uvt_1^n; H_{21}\ux_1^n + \uw_1^n + \uvt_1^n) \\
\stackrel{(e)} \leq & \ -nI(\uvt_{1}; H_{21}\ux_{1G} + \uw_1 + \uvt_1).
\end{split}
\end{equation}
where $\uv_1$ and $\uvt_1$ are Gaussian random vectors independent of every other random vector with the following covariance matrices:
\[
\begin{split}
\uv_1 \sim & \ \mathcal{N}\left(\uzero,I - A_2\SigmaGenie_2^{-1}A^{\top}_2\right) \\
\uvt_1 \sim & \ \mathcal{N}\left(\uzero,I - A_2\SigmaGenie_2^{-1}A^{\top}_2 - \SigmaGenie_1\right)
\end{split}
\]
Step (c) follows from Lemma~\ref{lemma6}, step (d) follows because $\uw_1 + \uvt_1$ has same distribution as $\uv_1$ and step (e) follows from Lemma~\ref{lemma4}.  Similarly,
\begin{equation} \label{eq:usefulStep3}
\begin{split}
\entropy{\us_2^n} - & \ \entropy{\uy_1^n|\us_1^n,\ux_1^n} \\
\leq & \ -nI(\uvt_{2}; H_{12}\ux_{2G} + \uw_2 + \uvt_2).
\end{split}
\end{equation}
The conditions in \eqref{eq:condition-useful} are required for the covariance matrices of $\uvt_1$ and $\uvt_2$ to be positive semi-definite. Substituting \eqref{eq:usefulStep2} and \eqref{eq:usefulStep3} in \eqref{eq:usefulStep1}, we get
\[
\begin{split}
\CsumGAIC \leq & \ \sum_{i = 1}^2 \entropy{\uy_{iG}|\us_{iG}} - \entropy{\uw_i}  \\ ~~~~~~~~~~~~~~~ & \ - I(\uvt_{1}; H_{21}\ux_{1G} + \uw_1 + \uvt_1) \\ ~~~~~~~~~~~~~~~ & \ - I(\uvt_{2}; H_{12}\ux_{2G} + \uw_2 + \uvt_2) \\
\end{split}
\]
Since all the inequalities in \eqref{eq:usefulStep1}, \eqref{eq:usefulStep2} and \eqref{eq:usefulStep3} are met equality when $\ux_i^n = \ux_{iG}^n$, we also have
\begin{equation} \label{eq:gaic-sumcap}
\begin{split}
\RGAIC(Q_1,Q_2) = & \ \sum_{i = 1}^2 I(\ux_{iG};\uy_{iG},\us_{iG}) \\
= & \ \sum_{i = 1}^2 \entropy{\uy_{iG}|\us_{iG}} - \entropy{\uw_i}  \\ ~~~~~~~~~~~~~~~ & \ - I(\uvt_{1}; H_{21}\ux_{1G} + \uw_1 + \uvt_1) \\ ~~~~~~~~~~~~~~~ & \ - I(\uvt_{2}; H_{12}\ux_{2G} + \uw_2 + \uvt_2) \\
\end{split}
\end{equation}
Therefore, we proved that
\[
\CsumGAIC \leq \RGAIC(Q_1,Q_2), \textrm{ for some }  Q_i \in \mathcal{Q}_i
\]
and hence
\[
\CsumGAIC \leq \max_{Q_i \in \mathcal{Q}_i}\RGAIC(Q_1,Q_2).
\]
We now prove that $\RGAIC(Q_1,Q_2)$ is concave in $(Q_1,Q_2)$ using the expression \eqref{eq:gaic-sumcap}. From Lemma~\ref{lemma:concavity}, it immediately follows that $\entropy{\uy_{1G}|\us_{1G}}$ and $\entropy{\uy_{2G}|\us_{2G}}$ are concave in $(Q_1,Q_2)$. Also, observe that $- I(\uvt_{2}; H_{12}\ux_{2G} + \uw_2 + \uvt_2)$ is also concave in $(Q_1,Q_2)$ since
\[
- I(\uvt_{2}; H_{12}\ux_{2G} + \uw_2 + \uvt_2) = -\entropy{\uvt_{2}} + \entropy{\uvt_{2}|H_{12}\ux_{2G} + \uw_2 + \uvt_2}
\]
where the first term does not depend on $(Q_1,Q_2)$ and Lemma~\ref{lemma:concavity} implies the concavity of the second term in $(Q_1,Q_2)$. Similarly, $- I(\uvt_{1}; H_{21}\ux_{1G} + \uw_1 + \uvt_1)$ is also concave in $(Q_1,Q_2)$.
\subsection{Proof of Lemma~\ref{lemma:mimo-smart}}
\label{app:mimo-smart}
\[
\begin{split}
\RGAIC(Q_1,Q_2,\Psi) = & \ \RIC(Q_1,Q_2) \\
\Leftrightarrow I(\ux_{iG};\uy_{iG},\us_{iG}) = & \ I(\ux_{iG};\uy_{iG}) \\
\Leftrightarrow I(\ux_{iG};\us_{iG}|\uy_{iG}) = & \ 0.
\end{split}
\]
Since $\ux_{1G}, \uy_{1G}$ and $\us_{1G}$ are all Gaussian, from \cite[Lemma~7]{Annapureddy-Veeravalli-IT-2008},
$I(\ux_{1G};\us_{1G}|\uy_{1G}) = \ 0$
iff the MMSE estimate of $\us_{1G}$ given $(\ux_{1G},\uy_{1G})$ is equal to the MMSE estimate of $\us_{1G}$ given $\uy_{1G}$. Using the property that the MMSE error is orthogonal to the observations $\ux_{1G}$ and $\uy_{1G}$, the above statement holds true iff there exist a matrix $T$ such that
\[
\begin{split}
\expect{(T\uy_{1G} - \us_{1G} )\ux_{1G}^{\top}} = & \ 0 \\
\expect{(T\uy_{1G} - \us_{1G} )\uy_{1G}^{\top}} = & \ 0
\end{split}
\]
Since $\uy_{1G} = H_{11}\ux_{1G} + H_{12}\ux_{2G} + \uz_1$, the above equations are true iff
\[
\begin{split}
\expect{(T\uy_{1G} - \us_{1G} )\ux_{1G}^{\top}} = & \ 0 \\
\expect{(T\uy_{1G} - \us_{1G} )(H_{12}\ux_{2G} + \uz_1)^{\top}} = & \ 0
\end{split}
\]
iff
\[
\begin{split}
(TH_{11} - H_{21})Q_1 = & \ 0 \\
T(H_{12}Q_2H_{12}^{\top} + I) - A_1^{\top}= & \ 0
\end{split}
\]
Solving for $T$ and substituting, we get
\[
\left(A_1^{\top}(H_{12}Q_2H_{12}^{\top} + I)^{-1}H_{11} - H_{21}\right)Q_1 = 0.
\]

\subsection{Sufficiency of two antennas for the two-user MISO and SIMO interference channels}
\label{app:twoAnts}
We argue that, without any loss of generality, we can assume two transmit antennas and
\[
\begin{split}
\uc = & \ [1 \ 0]^\top \\
\ud = & \ [\ct \ \st]^\top
\end{split}
\]
for the MISO interference channel \eqref{eq:miso-ic}. The argument for the sufficiency of two receive antennas for the SIMO interference channel follows in a similar fashion. Even though the argument is presented for the symmetric channel considered in this paper, the same applies for the asymmetric case as well.

Consider the vector $\ucp$ which is perpendicular to $\uc$ and is given by
\[
\ucp = \frac{\ud - \ct\uc}{\st}.
\]
If $\theta = 0$, use any vector perpendicular to $\uc$ as $\ucp$. Now let $U$ be an unitary matrix $\uc$ as its first column  $\ucp$
as its second column, and define $\tilde{\ux}_i$ as
\[
\tilde{\ux}_i = U^\top\ux_i
\]
or equivalently $\ux_i = U\tilde{\ux}_i$.

Therefore, the MISO interference channel \eqref{eq:miso-ic} can be re-written as
\[
\begin{split}
Y_1 = & \ \ud^\top{}U\tilde{\ux}_1 + h\uc^\top{}U\tilde{\ux}_2 + \uz_1 \\
		= & \ \tilde{\ud}^\top\tilde{\ux}_1 + h\tilde{\uc}^\top{}\tilde{\ux}_2 + \uz_1
\end{split}
\]
where
\[
\begin{split}
\tilde{\uc} = & \ [1 \ 0 \ \ 0 \cdots \ 0]^\top \\
\tilde{\ud} = & \ [\ct \ \st \ 0 \cdots \ 0]^\top
\end{split}
\]
The interference channel outputs $Y_1$ and $Y_2$ do not depend on channel inputs $\tilde{\ux}_{i3}$ to $\tilde{\ux}_{iM}$, $i = 1,2,$ and hence they can be dropped. Also, since $U$ is the unitary matrix, the average sum power constraint remains the same.
\subsection{Alternative expressions for MISO genie parameters}
\label{sec:miso_genie_expressions}
Let $J$ denote the matrix $I + h^2P\uc\uc^\top$. Therefore, from the matrix inversion Lemma, we have
\begin{equation} \label{eq:mat-inv}
\begin{split}
J^{-1}\ud = & \ \left(I - \frac{h^2P}{1 + h^2P}\uc\uc^\top\right)\ud = \ud - \frac{h^2P\ct}{1 + h^2P}\uc \\
\uc^{\top}J^{-1}\ud = & \ \ct - \frac{h^2P\ct}{1 + h^2P} = \frac{\ct}{1 + h^2P} \\
\ud^{\top}J^{-1}\ud = & \ 1 - \frac{h^2P\cst}{1 + h^2P} = \frac{1 + h^2P\sst}{1 + h^2P}.
\end{split}
\end{equation}
We will now provide alternative expressions for the parameters involved in MISO interference channel. These relations will be used in Appendices~\ref{sec:eqv-threshold} and \ref{sec:eigenValue}.
\begin{claim} \label{claim:b-expressions}
The beamforming vector $\ub$ defined in \eqref{eq:optBeam} can be expressed as
\begin{equation} \label{eq:exp-b}
\ub  = \frac{\ud}{||J^{-1}\ud||} - h^2P\ubc\uc.
\end{equation}
\end{claim}
\medskip
\begin{proof}
Using the notation $J = I + h^2P\uc\uc^\top$ in \eqref{eq:optBeam}, we get
\[
\begin{split}
\ub = & \ \frac{J^{-1}\ud}{||J^{-1}\ud||} \\
\end{split}
\]
Therefore, we have
\[
\begin{split}
J\ub = & \ \frac{\ud}{||J^{-1}\ud||} \\
\ub + h^2P\ubc\uc = & \ \frac{\ud}{||J^{-1}\ud||} \\
\ub = & \ \frac{\ud}{||J^{-1}\ud||} - h^2P\ubc\uc\\
\end{split}
\]
\end{proof}

\begin{claim} \label{claim:a-expressions}
The genie parameter $a$ defined in \eqref{eq:psi-miso} can be expressed as
\begin{equation} \label{eq:exp-a}
a = h\frac{\ubc}{\ubd}\ub^{\top}J\ub = h\frac{\ubc}{||J^{-1}\ud||} = h\frac{\uc^{\top}J^{-1}\ud}{\ud^{\top}J^{-2}\ud} = \frac{h(1+h^2P)\ct}{\cst + (1 + h^2P)^2\sst}.
\end{equation}
\end{claim}
\medskip
\begin{proof}
\[
\begin{split}
a = & \ h\frac{\ubc}{\ubd}\left(1 + h^2P|\ubc|^2\right) \\
   = & \ h\frac{\ubc}{\ubd}\ub^{\top}J\ub \\
	= & \ h\frac{\ubc}{||J^{-1}\ud||} \\
	= & \ h\frac{\uc^{\top}J^{-1}\ud}{||J^{-1}\ud||^2} \\
	= & \ h\frac{\uc^{\top}J^{-1}\ud}{\ud^{\top}J^{-2}\ud} \\
\end{split}
\]
which, using \eqref{eq:mat-inv}, can be further simplified as
\[
\begin{split}	
a	= & \ h\frac{\frac{\ct}{1 + h^2P}}{\frac{\cst}{(1 + h^2P)^2} + \sst} \\
  = & \ \frac{h(1+h^2P)\ct}{\cst + (1 + h^2P)^2\sst}
\end{split}
\]
\end{proof}

\begin{claim} \label{claim:mu-expressions}
The parameter $\mu$ defined in \eqref{eq:mu} can be expressed as
\begin{equation} \label{eq:exp-mu}
\mu = \frac{1 - \SigmaGenie + h^2P(\ub^{\top}\uc)^2}{a} = \frac{\ubd}{h\ubc} - \frac{\SigmaGenie}{a}
\end{equation}
\end{claim}
\begin{proof}
First, observe that
\[
\begin{split}
(\SigmaGenie + h^2P(\ub^{\top}\uc)^2)&(1 - \SigmaGenie + h^2P(\ub^{\top}\uc)^2) \\
 = & \ \SigmaGenie(1-\SigmaGenie) + h^2P(\ub^{\top}\uc)^2 + (h^2P(\ub^{\top}\uc)^2)^2 \\
 = & \ a^2 + h^2P(\ub^{\top}\uc)^2(1 + h^2P(\ub^{\top}\uc)^2) \\
 = & \ a^2 + hP\ubd\ubc \frac{h\ubc}{\ubd}(1 + h^2P(\ub^{\top}\uc)^2) \\
 = & \ a^2 + ahP\ubd\ubc \\
 = & \ a(a + hP\ubd\ubc). \\
\end{split}
\]
Using the above equality in \eqref{eq:mu}, we get
\[
\begin{split}
\mu = & \ \frac{a + hP\ub^{\top}\uc\ub^{\top}\ud}{\SigmaGenie + h^2P(\ub^{\top}\uc)^2} \\
		= & \ \frac{1 - \SigmaGenie + h^2P(\ub^{\top}\uc)^2}{a} \\
		= & \ \frac{\ub^{\top}J\ub - \SigmaGenie}{a} \\
		= & \ \frac{\ubd}{h\ubc} - \frac{\SigmaGenie}{a}.
\end{split}
\]
where the last step uses \eqref{eq:exp-a}.
\end{proof}
\subsection{Proof of \eqref{eq:thrcond1} and \eqref{eq:thrcond2}}
\label{sec:eqv-threshold}
\begin{lemma}
For any $P$, $h$ and $\theta \in \left(0,\frac{\pi}{2}\right)$ such that
\[
h < h_0(\theta,P)
\]
where $h_0(\theta,P)$, as defined in \eqref{eq:condition-simo}, is the positive solution to the equation
\[
h_0^2 - \sst = \left(\frac{\ct}{1 + h_0^2P} - h_0 \right)_+^2
\]
the following conditions are satisfied:
\begin{equation} \label{eq:miso_cond}
\begin{split}
a < & \ 0.5 \\
\mu{}h < & \ \ct
\end{split}
\end{equation}
where $a$ and $\mu$ are as defined in \eqref{eq:psi-miso} and \eqref{eq:mu} respectively.
\end{lemma}
\medskip
\begin{proof}
First, observe that $h < h_0(\theta,P)$ implies
\begin{equation} \label{eq:temp10}
h^2 - \sst < h_0^2 - \sst = \left(\frac{\ct}{1 + h_0^2P} - h_0 \right)_+^2 < \left(\frac{\ct}{1 + h^2P} - h \right)_+^2
\end{equation}
This immediately implies that
\[
\begin{split}
h^2 - \sst < \left(\frac{\ct}{1 + h^2P} - h \right)^2
\end{split}
\]
which is equivalent to 
\[
\begin{split}
\frac{\cst}{(1+h^{2}P)^{2}} - \frac{2h\ct}{1+h^{2}P} + \sst & \ > 0 \\
\cst -2h(1+h^{2}P)\ct + (1+h^{2}P)\sst & \ > 0 \\
2h(1+h^{2}P)\ct & \ < \cst + (1+h^{2}P)\sst \\
\Rightarrow a = \frac{2h(1+h^{2}P)\ct}{\cst + (1+h^{2}P)\sst} & \ < 0.5
\end{split}
\]
where we used the expression for $a$ from Claim~\ref{claim:a-expressions} in Appendix~\ref{sec:miso_genie_expressions}. Now it remains to prove that $\mu{}h < \ct$ is also satisfied. From Claim~\ref{claim:mu-expressions} in Appendix~\ref{sec:miso_genie_expressions}, we have
\[
\mu = \frac{\ubd}{h\ubc} - \frac{\SigmaGenie}{a}
\]
and thus
\[
\begin{split}
\mu{}h - \ct = & \ \frac{\ubd}{\ubc} - \frac{h\SigmaGenie}{a} - \ct\\
						 = & \ \frac{\ud^{\top}J^{-1}\ud}{\uc^{\top}J^{-1}\ud} - \frac{h\SigmaGenie}{a} - \ct \\
						 = & \ \frac{1 + h^2P\sst}{\ct} - \frac{h\SigmaGenie}{a} - \ct \\
						 = & \ \frac{(1 + h^2P)\sst}{\ct} - \frac{h\SigmaGenie}{a}
\end{split}
\]
where $J$ is used to denote the matrix $I + h^{2}P\uc\uc^{\top}$. If $h$ satisfies \eqref{eq:temp10}, then one of the following cases has to be true.\\
\textbf{Case 1:} $h$ satisfies
\[
h < \st.
\]
Therefore, we have
\[
\begin{split}
\frac{h\SigmaGenie}{a} = & \ h\frac{0.5 + 0.5\sqrt{1 - 4a^2}}{a} \\
= & \ \frac{h}{2}\left(\frac{1}{a} + \sqrt{\frac{1}{a^2}-4}\right) \\
\stackrel{(a)} = & \ \frac{h}{2}\left(\frac{\cst + (1 + h^2P)^2\sst}{h(1+h^2P)\ct} + \sqrt{\left(\frac{\cst + (1 + h^2P)^2\sst}{h(1+h^2P)\ct}\right)^2 -4}\right) \\
= & \ \frac{1}{2(1+h^2P)\ct}\left(\cst + (1 + h^2P)^2\sst + \sqrt{\left(\cst + (1 + h^2P)^2\sst\right)^2 - 4h^2(1+h^2P)^2\ct}\right) \\
\stackrel{(b)} > & \ \frac{1}{2(1+h^2P)\ct}\left(\cst + (1 + h^2P)^2\sst + \sqrt{\left(\cst + (1 + h^2P)^2\sst\right)^2 - 4\sst(1+h^2P)^2\cst}\right) \\
= & \ \frac{1}{2(1+h^2P)\ct}\left(\cst + (1 + h^2P)^2\sst + \sqrt{\left(\cst - (1 + h^2P)^2\sst\right)^2}\right) \\
\geq & \ \frac{1}{2(1+h^2P)\ct}\left(\cst + (1 + h^2P)^2\sst - \cst + (1 + h^2P)^2\sst\right) \\
= & \ \frac{(1 + h^2P)\sst}{\ct} \\
\end{split}
\]
where step (a) uses an expression for $a$ from Claim~\ref{claim:a-expressions} in Appendix~\ref{sec:miso_genie_expressions}, and step (b) follows because $h < \st$. Therefore,
\[
\mu{}h - \ct = \frac{(1 + h^2P)\sst}{\ct} - \frac{h\SigmaGenie}{a} < 0.
\]
\textbf{Case 2:} $h$ satisfies
\begin{equation} \label{eq:temp11}
\begin{split}
h \geq & \ \st \\
\frac{\ct}{1 + h^2P} - h  \geq & \ 0.
\end{split}
\end{equation}
Note that
\[
\SigmaGenie = 0.5 + 0.5\sqrt{1 - a^2} > 0.5 > a.
\]
Hence, we have
\[
\begin{split}
\mu{}h - \ct      = & \ \frac{(1 + h^2P)\sst}{\ct} - \frac{h\SigmaGenie}{a} \\
< & \ \frac{(1 + h^2P)\sst}{\ct} - h \\
\stackrel{(a)} \leq & \ \frac{(1 + h^2P)\sst}{\ct} - \frac{\sst}{h} \\
					        = & \ \sst\left(\frac{1 + h^2P}{\ct} - \frac{1}{h}\right) \\
\stackrel{(b)} \leq & \ 0
\end{split}
\]
where steps (a) and (b) follow from \eqref{eq:temp11}.
\end{proof}
\subsection{Proof of \eqref{eq:step2}}
\label{sec:eigenValue}
In this Appendix, we show that if
\begin{equation} \label{eq:muh-lessthan-cos}
\mu{}h < \cos\theta
\end{equation}
then $\tr{MQ} \leq \tr{MQ^*}, \forall Q \in \mathcal{Q}$, where $\mu$ and $M$ are defined in \eqref{eq:mu} and \eqref{eq:M} respectively. For all $Q \in \mathcal{Q}$, we have
\begin{equation}
\begin{split}
\tr{MQ} \leq & \ \tr{Q}\lambdaMax(M) \\
				\leq & \ P\lambdaMax(M) \\
\stackrel{(b)} = & \ \ P\ub^{\top}M\ub \\
				= & \ \tr{MP\ub\ub^{\top}} \\
				= & \ \tr{MQ^*}.
\end{split}
\end{equation}
Step (b) follows because $\ub$ is the eigenvector of the matrix $M$ corresponding to the maximum eigenvalue $\lambdaMax(M)$. In proving that $\ub$ is an eigenvector of the matrix $M$, we need the following relation.
\begin{claim} \label{claim:appG1}
\begin{equation} \label{eq:rel2}
\left(\ubd - \mu{}h\ubc\right)\mu{h} - h^2\ubc = h^2P\ubc\left(\ubd - \mu{}h\ubc\right)||J^{-1}\ud||
\end{equation}
\end{claim}
\medskip
\begin{proof}
Using the two different expressions for $\mu$ in Claim~\ref{claim:mu-expressions} in Appendix~\ref{sec:miso_genie_expressions}, we get
\[
\begin{split}
\left(\frac{\ubd}{h\ubc} - \mu\right)\left(\mu - \frac{h^2P(\ub^{\top}\uc)^2}{a}\right) = & \ \frac{\SigmaGenie(1-\SigmaGenie)}{a^2} = 1 \\
\end{split}
\]
Using \eqref{eq:exp-a}, we get
\[
\begin{split}
\left(\frac{\ubd}{h\ubc} - \mu\right)\left(\mu - P||J^{-1}\ud||h\ubc\right) = & \ 1 \\
\left(\ubd - \mu{}h\ubc\right)\left(\mu{}h - ||J^{-1}\ud||h^2P\ubc\right) = & \ h^2\ubc.
\end{split}
\]
\end{proof}
We now prove that $\ub$ is an eigenvector of the matrix $M$:
\[
\begin{split}
M\ub = & \ \left((\ud - \mu{}h\uc)(\ud - \mu{}h\uc)^{\top} + h^2\uc\uc^{\top}\right)\ub \\
		 = & \ (\ubd - \mu{}h\ubc)(\ud - \mu{}h\uc) + h^2\ubc\uc \\
		 = & \ (\ubd - \mu{}h\ubc)\ud - ((\ubd - \mu{}h\ubc)\mu{}h - h^2\ubc)\uc \\
\stackrel{(a)} = & \ (\ubd - \mu{}h\ubc)\ud - h^2P\ubc\left(\ubd - \mu{}h\ubc\right)||J^{-1}\ud||\uc \\
		 = & \ \left(\ubd - \mu{}h\ubc\right)||J^{-1}\ud||\left(\frac{\ud}{||J^{-1}\ud||} - h^2P\ubc\uc\right) \\
\stackrel{(b)} = & \ \left(\ubd - \mu{}h\ubc\right)||J^{-1}\ud||\ub.
\end{split}
\]
where step (a) follows from Claim~\ref{claim:appG1}, and step (b) follows from Claim~\ref{claim:b-expressions} in Appendix~\ref{sec:miso_genie_expressions}. From Lemma~\ref{lemma:eig}, it follows that $\ub$ corresponds to the largest eigenvalue of the matrix $M = (\ud - \mu{}h\uc)(\ud - \mu{}h\uc)^{\top} + h^2\uc\uc^{\top}$. The conditions required in Lemma~\ref{lemma:eig} are shown below.
\[
\begin{split}
(\ud - \mu{}h\uc)^\top\uc = & \ \ct - \mu{}h \\
\stackrel{(a)}						> & \ 0
\end{split}
\]
Also,
\[
\begin{split}
\ub^{\top}\uc = & \ \frac{\uc^{\top}J^{-1}\ud}{||J^{-1}\ud||} \\
\stackrel{(b)}= & \ \frac{\ct}{(1 + h^2P)||J^{-1}\ud||} \\
\stackrel{(c)}> & \ 0
\end{split}
\]
and
\[
\begin{split}
(\ud - \mu{}h\uc)^\top\ub  = & \ \frac{(\ud - \mu{}h\uc)^{\top}J^{-1}\ud}{||J^{-1}\ud||} \\
\stackrel{(d)}	 = & \ \frac{1 + h^2P\sst - \mu{}h\ct}{(1 + h^2P)||J^{-1}\ud||} \\
\stackrel{(e)}	 > & \ \frac{1 + h^2P\sst - \cst}{(1 + h^2P)||J^{-1}\ud||} \\
= & \ \frac{\sst + h^2P\sst}{(1 + h^2P)||J^{-1}\ud||} \\
= & \ \frac{\sst}{||J^{-1}\ud||} \\
\stackrel{(f)} 	> & \ 0
\end{split}
\]
where steps (a) and (e) follow from the hypothesis \eqref{eq:muh-lessthan-cos}, steps (b) and (d) follow from \eqref{eq:mat-inv}, steps (c) and (f) follow because of our assumption that $\theta \in \left(0,\frac{\pi}{2}\right)$.
\subsection{Proof of Lemma~\ref{lemma:simo-useful}}
\label{app:simo-useful}
The proof is very similar to the proof of Lemma~\ref{lemma:mimo-useful} in Appendix~\ref{app:mimo-useful}. The only difference is steps \eqref{eq:usefulStep2} and \eqref{eq:usefulStep3}, where we showed that $\entropy{\us_1^n} - \entropy{\uy_2^n|\us_2^n,\ux_2^n}$ and $\entropy{\us_2^n} - \entropy{\uy_1^n|\us_1^n,\ux_1^n}$ are maximized by $X_{1G}^n$ and $X_{2G}^n$, respectively. Here we show that the condition \eqref{eq:condition-useful} is not necessary and a weaker condition:
\[
\inv{\uc^{\top}\SigmaGenie^{-1}\uc} \leq \inv{\uc^{\top}\inv{I - A\SigmaGenie^{-1}A^{\top}}\uc}
\]
is sufficient for \eqref{eq:usefulStep2} and \eqref{eq:usefulStep3} to be true. We give the proof assuming $\uc = [1 \ 0]^\top$ to simplify the presentation, but can be extended to arbitrary $\uc$ using the arguments in Appendix~\ref{app:twoAnts}. Following \eqref{eq:usefulStep2}, we have
\[
\entropy{\us_1^n} - \entropy{\uy_2^n|\us_2^n,\ux_2^n} = \entropy{h\uc{}X_1^n + \uw_1^n } - \entropy{h\uc{}X_1^n + \uv_1^n}
\]
where $\uw_1 \sim \mathcal{N}(\uzero,\SigmaGenie)$ and $\uv_1 \sim \mathcal{N}(\uzero,I - A\SigmaGenie^{-1}A^{\top})$. Now,
\[
\begin{split}
\entropy{h\uc{}X_1^n + \uw_1^n } = & \ \entropy{hX_1^n + W_{11}^n, W_{12}^n }  \\
																 = & \ \entropy{hX_1^n + W_{11}^n|W_{12}^n }  + \entropy{W_{12}^n }  \\
																 = & \ \entropy{hX_1^n + \hat{W}_1^n}  + n\entropy{W_{12}}  \\
\end{split}
\]
where $\hat{W}_1 \sim \mathcal{N}(0,\cov{W_{11}|W_{12}})$. Let
\[
\SigmaGenie = \cov{\uw_1} = \left[ \begin{array}{c c} a_1 & b \\ b & a_2 \end{array}\right]
\]
then
\[
\SigmaGenie^{-1} = \frac{1}{a_1 a_2 - b^2}\left[ \begin{array}{c c} a_2 & -b \\ -b & a_1 \end{array}\right]
\]
and since $\uc = [1 \ 0]^\top$, we have
\[
\begin{split}
\inv{\uc^{\top}\SigmaGenie^{-1}\uc} = & \ \frac{a_1 a_2 - b^2}{a_2} \\
= & \ a_1 - \frac{b^2}{a_2} \\
= & \ \cov{W_{11}|W_{12}} = \cov{\hat{W}_1}.
\end{split}
\]
Similarly,
\[
\entropy{h\uc{}X_1^n + \uv_1^n } = \entropy{hX_1^n + \hat{V}_1^n}  + n\entropy{V_{12}}
\]
where
\[
\hat{V}_1 \sim \mathcal{N}\left(0,\inv{\uc^{\top}\inv{I - A\SigmaGenie^{-1}A^{\top}}\uc}\right)
\]
Therefore,
\[
\begin{split}
\entropy{\us_1^n} - \entropy{\uy_2^n|\us_2^n,X_2^n} = & \ \entropy{h\uc{}X_1^n + \uw_1^n } - \entropy{h\uc{}X_1^n + \uv_1^n} \\
= & \ \entropy{hX_1^n + \hat{W}_1^n}  - \entropy{hX_1^n + \hat{V}_1^n} + n\entropy{W_{12}}   -  n\entropy{V_{12}} \\
\leq & \ n\entropy{hX_{1G} + \hat{W}_1}  - n\entropy{hX_{1G} + \hat{V}_1} + n\entropy{W_{12}}   -  n\entropy{V_{12}}.
\end{split}
\]
We have thus shown that $\entropy{\us_1^n} - \entropy{\uy_2^n|\us_2^n,X_2^n}$ is maximized by $X_{1G}^n$ and similarly we can show that $\entropy{\us_2^n} - \entropy{\uy_1^n|\us_1^n,X_1^n}$ is maximized by $X_{2G}^n$.
\subsection{Existence of SIMO genie parameters}
\label{app:simo}
Here we show that if $h \leq h_0(\theta,P)$ where $h_0(\theta,P)$ is the positive solution of
\[
h^2 - \sst = \left(\frac{\ct}{1 + h^2P} - h \right)_+^2.
\]
then there exist $k$, $\eta$ and $\vv$ satisfying the conditions
\begin{equation} \label{eq:conds-simo}
\begin{split}
\eta \geq & \ k^2 \\
\eta \leq & \ \frac{\eta - k^2}{\eta - k^2(1 - (\ucv)^2)} \\
h = & \ k\vv^\top\inv{I + h^2P\uc\uc^{\top}}\ud.
\end{split}
\end{equation}
First, we eliminate $\eta$, and obtain conditions on $k$ and $\vv$. Let $t = \eta - k^{2}$, then the second condition in \eqref{eq:conds-simo} is equivalent to
\[
\begin{split}
t + k^{2} \leq \frac{t}{t + k^{2}(\uc^{\top}\vv)^{2}} \\
\Rightarrow t^{2} + t(k^{2}\left(1 + (\uc^{\top}\vv)^{2}) - 1\right) + k^{4}(\uc^{\top}\vv)^{2} \leq 0.
\end{split}
\]
The following claim can be easily verified.
\begin{claim}
A non-negative solution ($t \geq 0$) to the inequality $t^{2} + bt + c \leq 0,$ with $c \geq 0$, exists iff $b \leq -2\sqrt{c}$.
\end{claim}
Applying the above claim, a valid $t$ exists iff $k$ and $\vv$ satisfy
\[
\begin{split}
k^{2}(1 + (\uc^{\top}\vv)^{2}) - 1 \leq - 2|\uc^{\top}\vv| \\
\Rightarrow k^{2}(1 + |\uc^{\top}\vv|)^{2} \leq 1 \\
\Rightarrow k(1 + |\uc^{\top}\vv|) \leq 1.
\end{split}
\]
Now, using the last condition in \eqref{eq:conds-simo} and eliminating $k$, we get
\[
h(1 + |\uc^{\top}\vv|) \leq \vv^{\top}\inv{I + h^2P\uc\uc^{\top}}\ud.
\]
With out any loss of generality, let $\vv = [\alpha \ \beta]^{\top}$ such that $\alpha^{2} + \beta^{2} = 1$. Using the assumptions that $\uc = [1 \ 0]^{\top}$ and $\ud = [\ct \ \st]^{\top}$, the above inequality is equivalent to
\[
h(1 + |\alpha|) \leq \alpha\frac{\ct}{1 + h^{2}P} + \beta\st.
\]
 Observe that we can restrict ourselves to only positive values for $\alpha$, because if a negative $\alpha$ works then the corresponding positive $\alpha$ works as well. Thus, we reduced the problem to finding the conditions on $h$, $\theta$ and $P$, so that there exist $\alpha > 0$ and $\beta$ exist satisfying $\alpha^{2} + \beta^{2} = 1$
\[
\begin{split}
h \leq \alpha\left(\frac{\ct}{1+h^{2}P} - h\right) + \beta\st.
\end{split}
\]
Now, observe that if $\frac{\ct}{1+h^{2}P} - h < 0$, then the best is to set $\alpha = 0$ and $\beta = 1$, giving rise to the condition 
\[
h \leq \st.
\]
Otherwise, if $\frac{\ct}{1+h^{2}P} - h \geq 0$, then the best is choose $\alpha$ and $\beta$ proportional to $\frac{\ct}{1+h^{2}P} - h$ and $\st$ respectively, giving rise to the condition
\[
h^{2} \leq \left(\frac{\ct}{1+h^{2}P} - h\right)^{2} + \sst.
\]
Thus, we proved that the genie parameters $\eta,k,\vv$ exist satisfying \eqref{eq:conds-simo} iff $h,\theta$ and $P$ satisfy
\[
h^{2}  - \sst \leq \left(\frac{\ct}{1+h^{2}P} - h\right)_{+}^{2}.
\]
It is easy to verify that the above is a threshold condition on $h$: $h \leq h_{0}(\theta,P)$. 
\bibliographystyle{IEEE}
\bibliography{MIMOIC}

\begin{thebibliography}{10}

\bibitem{Carleial1975}
A.~B. Carleial,
\newblock ``{A case where interference does not reduce capacity}'',
\newblock {\em IEEE Trans. Inform. Th.}, vol. 21, no. 1, pp. 569--570, Sept.
  1975.

\bibitem{HK1981}
T.~S. Han and K.~Kobayashi,
\newblock ``{A new achievable rate region for the interference channel}'',
\newblock {\em IEEE Trans. Inform. Th.}, vol. 27, no. 1, pp. 49--60, Jan. 1981.

\bibitem{Sato1981}
H.~Sato,
\newblock ``{The capacity of the Gaussian interference channel under strong
  interference}'',
\newblock {\em IEEE Trans. Inform. Th.}, vol. 27, no. 6, pp. 786--788, Nov.
  1981.

\bibitem{OneBit2008}
R.~H. Etkin, D.~N.~C. Tse, and H.~Wang,
\newblock ``{Gaussian interference channel capacity to within one bit}'',
\newblock {\em IEEE Trans. Inform. Th.}, vol. 54, no. 12, pp. 5534--5562, Dec.
  2008.

\bibitem{Shang-Kramer-Chen-IT-2008}
X.~Shang, G.~Kramer, and B.~Chen,
\newblock ``{A new outer bound and noisy-interference sum-rate capacity for the
  Gaussian interference channels}'',
\newblock {\em IEEE Trans. Inform. Th.}, vol. 55, no. 2, pp. 689--699, Feb.
  2009.

\bibitem{Motahari-Khandani-IT-2008}
A.~S. Motahari and A.~K. Khandani,
\newblock ``{Capacity bounds for the Gaussian interference channel}'',
\newblock {\em IEEE Trans. Inform. Th.}, vol. 55, no. 2, pp. 620--643, Feb.
  2009.

\bibitem{Annapureddy-Veeravalli-IT-2008}
V.~S. Annapureddy and V.~V. Veeravalli,
\newblock ``Gaussian interference networks: {S}um capacity inthe low
  interference regime and new outerbounds on the capacity region'',
\newblock {\em IEEE Trans. Inform. Th.}, vol. 55, no. 6, pp. 3032--3050, June
  2009.

\bibitem{Vishwanath-Jafar-04}
S.~Vishwanath and {S.A.} Jafar,
\newblock ``On the capacity of vector gaussian interference channels'',
\newblock in {\em Information Theory Workshop, 2004. {IEEE}}, 2004, pp.
  365--369.

\bibitem{Shang-Chen-Gans-06}
Xiaohu Shang, Biao Chen, and {M.J.} Gans,
\newblock ``On the achievable sum rate for {MIMO} interference channels'',
\newblock {\em IEEE Trans. Inform. Th.}, vol. 52, no. 9, pp. 4313--4320, 2006.

\bibitem{TelatarTse2007}
E.~Telatar and D.~Tse,
\newblock ``{Bounds on the capacity region of a class of interference
  channels}'',
\newblock in {\em Proc. International Symposium on Information Theory, Nice,
  France}, June 2007.

\bibitem{Shang-et-al-Allerton2008}
X.~Shang, B.~Chen, G.~Kramer, and H.~V. Poor,
\newblock ``{On the Capacity of MIMO Interference Channels}'',
\newblock in {\em Proceedings of 46th Annual Allerton Conf. Commun. Cont. and
  Comp., University of Illinois, IL}, Oct. 2008.

\bibitem{shang_capacity_2009}
Xiaohu Shang, Biao Chen, Gerhard Kramer, and H.~Vincent Poor,
\newblock ``Capacity regions and {Sum-Rate} capacities of vector gaussian
  interference channels'',
\newblock {\em Submitted to IEEE Trans. Inform. Th.}, July 2009.

\bibitem{Bernd-etal-Asilomar2008}
Bernd Bandemer, Aydin Sezgin, and Arogyaswami Paulraj,
\newblock ``On the noisy interference regime of the {MISO} gaussian
  interference channel'',
\newblock in {\em Signals, Systems and Computers, 2008 42nd Asilomar Conference
  on}, 2008, pp. 1098--1102.

\bibitem{shang_multi-user_2009}
Xiaohu Shang, Biao Chen, and H.~Vincent Poor,
\newblock ``{Multi-User} {MISO} interference channels with {Single-User}
  detection: Optimality of beamforming and the achievable rate region'',
\newblock {\em Submitted to IEEE Trans. Inform. Th.}, July 2009.

\bibitem{horn-jhonson}
Roger Horn and Charles~R. Johnson,
\newblock {\em Matrix Analysis},
\newblock Cambridge University Press, New York, 1985.

\bibitem{boyd-convex-optimization}
Stephen Boyd and Lieven Vandenberghe,
\newblock {\em {Convex optimization}},
\newblock Cambridge University Press, New York, 2006.

\bibitem{jorswieck_complete_2008}
{E.A.} Jorswieck, {E.G.} Larsson, and D.~Danev,
\newblock ``Complete characterization of the pareto boundary for the {MISO}
  interference channel'',
\newblock {\em Signal Processing, {IEEE} Transactions on}, vol. 56, no. 10, pp.
  5292--5296, 2008.

\bibitem{zakhour}
R.~Zakhour and D.~Gesbert,
\newblock ``{Coordination on the MISO Interference Channel using the Virtual
  SINR Framework}'',
\newblock in {\em Proc. International ITG Workshop on Smart Antennas}, 2009.

\bibitem{Thomas-IT1987}
J.~A. Thomas,
\newblock ``{Feedback can at most double Gaussian multiple access channel
  capacity}'',
\newblock {\em IEEE Trans. Inform. Th.}, vol. 33, no. 5, pp. 711--716, Sept.
  1987.

\end{thebibliography}
\end{document}